\documentclass[aps,prl,twocolumn,superscriptaddress,showpacs,longbibliography,floatfix]{revtex4-2}

\usepackage{amssymb}
\usepackage{graphicx}
\usepackage{amsmath}
\usepackage{mathtools}
\usepackage[export]{adjustbox}
\usepackage{epsfig}
\usepackage{times}
\usepackage{color}
\usepackage{subfigure}
\usepackage{setspace}
\usepackage{natbib}
\usepackage[normalem]{ulem}
\usepackage{cancel}
\usepackage{soul}
\usepackage{physics}

\usepackage[pdftex]{hyperref}
\usepackage{ulem} 
\hypersetup{colorlinks = true, urlcolor = blue, linkcolor = blue, citecolor = blue}

\usepackage{xcolor}
\usepackage{xparse,xcoffins}
\ExplSyntaxOn
\NewCoffin\imagecoffin
\NewCoffin\labelcoffin
\keys_define:nn { miguel/label }
 {
  label   .tl_set:N = \l_miguel_label_tl,
  labelbox .bool_set:N = \l_miguel_label_box_bool,
  labelbox .default:n = true,
  fontsize .tl_set:N = \l_miguel_label_size_tl,
  fontsize .initial:n = \footnotesize,
  pos .choice:,
  pos/nw .code:n = \tl_set:Nn \l_miguel_label_pos_tl { left,up },
  pos/ne .code:n = \tl_set:Nn \l_miguel_label_pos_tl { right,up },
  pos/sw .code:n = \tl_set:Nn \l_miguel_label_pos_tl { left,down },
  pos/se .code:n = \tl_set:Nn \l_miguel_label_pos_tl { right,down },
  pos/n .code:n = \tl_set:Nn \l_miguel_label_pos_tl { hc,up },
  pos/w .code:n = \tl_set:Nn \l_miguel_label_pos_tl { left,vc },
  pos/s .code:n = \tl_set:Nn \l_miguel_label_pos_tl { hc,down },
  pos/e .code:n = \tl_set:Nn \l_miguel_label_pos_tl { right,vc },
  pos .initial:n = nw,
  unknown .code:n   = \clist_put_right:Nx \l_miguel_label_clist
                       { \l_keys_key_tl = \exp_not:n { #1 } }
 }
\clist_new:N \l_miguel_label_clist
\box_new:N \l_miguel_label_box
\box_new:N \l_miguel_label_image_box
\NewDocumentCommand{\xincludegraphics}{O{}m}
 {
  \group_begin:
  \tl_clear:N \l_miguel_label_tl
  \clist_clear:N \l_miguel_label_clist
  \keys_set:nn { miguel/label } { #1 }
  \tl_if_empty:NTF \l_miguel_label_tl
   {
    \miguel_includegraphics:Vn \l_miguel_label_clist { #2 }
   }
   {
    \SetHorizontalCoffin\imagecoffin
     {
      \miguel_includegraphics:Vn \l_miguel_label_clist { #2 }
     }
    \SetHorizontalCoffin\labelcoffin
     {
      \raisebox{\depth}
       {
        \bool_if:NTF \l_miguel_label_box_bool
         { \fcolorbox{white}{white}{\l_miguel_label_size_tl\l_miguel_label_tl} }
         { \l_miguel_label_size_tl\l_miguel_label_tl }
       }
     }
    \SetVerticalPole\imagecoffin{left}{3pt+\CoffinWidth\labelcoffin/2}
    \SetVerticalPole\imagecoffin{right}{\Width-3pt-\CoffinWidth\labelcoffin/2}
    \SetHorizontalPole\imagecoffin{up}{\Height-3pt-\CoffinHeight\labelcoffin/2}
    \SetHorizontalPole\imagecoffin{down}{3pt+\CoffinHeight\labelcoffin/2}
    \use:x{\JoinCoffins\imagecoffin[\l_miguel_label_pos_tl]\labelcoffin[vc,hc]} 
    \TypesetCoffin\imagecoffin
   }
   \group_end:
 }
\NewDocumentCommand{\setlabel}{m}
 {
  \keys_set:nn { miguel/label } { #1 }
 }
\cs_new_protected:Nn \miguel_includegraphics:nn
 {
  \includegraphics[#1]{#2}
 }
\cs_generate_variant:Nn \miguel_includegraphics:nn { V }
\ExplSyntaxOff

\begin{document}

\begin{abstract}
The quantum metric of single-particle wave functions in topological flatbands plays a crucial role in determining the stability of fractional Chern insulating (FCI) states. Here, we unravel that the quantum metric causes the many-body Chern number of the FCI states to deviate sharply from the expected value associated with partial filling of the single-particle topological flatband. Furthermore, the variation of the quantum metric in momentum space induces band dispersion through interactions, affecting the stability of the FCI states. This causes a reentrant transition into the Fermi liquid from the FCI phase as the interaction strength increases. 
\end{abstract}

\title{Quantum-metric-induced quantum Hall conductance inversion and reentrant transition in fractional Chern insulators}
\author{Ang-Kun Wu}
\affiliation{Theoretical Division, T-4, Los Alamos National Laboratory (LANL), Los Alamos, New Mexico 87545, USA}
\author{Siddhartha Sarkar}
\affiliation{Department of Physics, University of Michigan, Ann Arbor, Michigan, 48109, USA}
\author{Xiaohan Wan}
\affiliation{Department of Physics, University of Michigan, Ann Arbor, Michigan, 48109, USA}
\author{Kai Sun}
\email{sunkai@umich.edu}
\affiliation{Department of Physics, University of Michigan, Ann Arbor, Michigan, 48109, USA}
\author{Shi-Zeng Lin}
\email{szl@lanl.gov}
\affiliation{Theoretical Division, T-4 and CNLS, Los Alamos National Laboratory (LANL),
Los Alamos, New Mexico 87545, USA}
\affiliation{Center for Integrated Nanotechnology, Los Alamos National Laboratory (LANL),
Los Alamos, New Mexico 87545, USA}
\date{\today}

\maketitle


 \textit{Introduction.} The interplay between strong correlation and topology underpins many emergent phenomena in condensed matter systems. A seminal example of this is the discovery of the fractional quantum Hall effect (FQHE) in electron gases subjected to a strong perpendicular magnetic field, forming Landau levels \cite{PhysRevLett.48.1559,RevModPhys.71.875,PhysRevLett.50.1395,PhysRevLett.90.016801}. A key feature of the FQHE is the fractionally quantized Hall conductance, which correlates with the Chern number of the Landau level and the fractional filling \cite{PhysRevLett.50.1395,trivedi1991numerical,PhysRevLett.63.199,jain2007composite,PhysRevB.97.035149}. The recent discovery of the fractional Chern insulator (FCI) under zero magnetic field in twisted MoTe\textsubscript{2}  moir\'e superlattice \cite{park2023observation,cai2023signatures,zeng2023thermodynamic,kang2024evidence} and pentalayer graphene \cite{lu2024fractional,Xie_Huo_Lu_2024} highlights the profound interplay between strong correlation and topology in band systems. In twisted MoTe\textsubscript{2}, the FCI appears through the partial filling of a topological flat band, with conductance determined by the single-particle band topology \cite{PhysRevResearch.3.L032070,PhysRevB.107.L201109}. In contrast, the emergence of FCI in pentalayer graphene is unexpected, since the stabilization of a Chern band itself necessitates interaction.

The behavior of electrons in quantum materials is governed by the energy dispersion and their wave functions in the Hilbert space. An important characterization of the structure of the wave function is the quantum geometric tensor, $\eta_{\mu\nu}(\mathbf{k})\equiv \langle \partial_\mu u_\mathbf{k}|(1-|u_\mathbf{k}\rangle\langle u_\mathbf{k}|)|\partial_\nu u_\mathbf{k}\rangle$, where $\ket{u_\mathbf{k}}$ is the periodic part of the Bloch wavefunction \cite{provost1980riemannian}. The real part of $\eta_{\mu\nu}$ is the Fubini-Study quantum metric, $g_{\mu\nu}=\mathrm{Re}[\eta_{\mu\nu}]$, and the imaginary part is the Berry curvature, $F_{xy}=-2\mathrm{Im}[\eta_{xy}]$, which determines the topology of quantum systems.
The role of topology in quantum systems has been well recognized, as demonstrated in the integer/fractional quantum Hall effect \cite{PhysRevLett.45.494,PhysRevLett.49.405,PhysRevLett.48.1559,PhysRevLett.50.1395}, Chern insulators \cite{PhysRevLett.61.2015}, and more recently topological insulators \cite{PhysRevLett.95.146802} and semimetals \cite{PhysRevB.83.205101}. The quantum metric, which is another important aspect of a quantum system, has attracted considerable attention only very recently. Authors of recent studies have shown that quantum geometry can induce transport phenomena \cite{PhysRevLett.127.277201,PhysRevLett.112.166601,PhysRevLett.132.026301,Gao_Liu_Qiu_Ghosh2023,PhysRevLett.122.227402} and can be crucial for the stability of quantum states \cite{peotta2015superfluidity,torma2022superconductivity}, including FCI \cite{parameswaran2013fractional,Bergholtz_Liu_2013,neupert2015fractional,LIU2024515,PhysRevResearch.5.L032022,PhysRevLett.132.096602,PhysRevB.110.035130,shavit2024quantum}. The stabilization of FCI hinges on several factors, such as band flatness, uniform Berry curvature distribution across momentum space, and the trace condition, which connects the real and imaginary parts of quantum geometry. Theoretically \cite{PhysRevLett.106.236802,PhysRevLett.106.236803,PhysRevLett.106.236804,sheng2011fractional,PhysRevX.1.021014,PhysRevResearch.3.L032070,PhysRevLett.131.136502, PhysRevA.108.032218}, the approach to achieve an FCI state starts with a flat Chern band, followed by optimizing the quantum geometry of the single-particle wave function \cite{PhysRevB.90.165139,PhysRevResearch.2.023237,PhysRevLett.127.246403,PhysRevResearch.5.023167,PhysRevB.33.2481,PhysRevB.103.205413,PhysRevB.104.045103,PhysRevB.104.045104,PhysRevB.104.115160,PhysRevB.108.205144,fujimoto2024higher}. 
Then partially filling the flat topological band stabilizes the FCI with Hall conductance $\sigma_{xy}=C\nu e^2/h$, where $C$ is the Chern number of the band and $\nu$ is the filling factor. However, the discovery of FCI in pentalayer graphene challenges this paradigm and calls for a scrutiny of the relationship between the single-particle band and the many-body quantum state.

\begin{figure}[t!]
\begin{center}
\includegraphics[width = 0.49\textwidth]{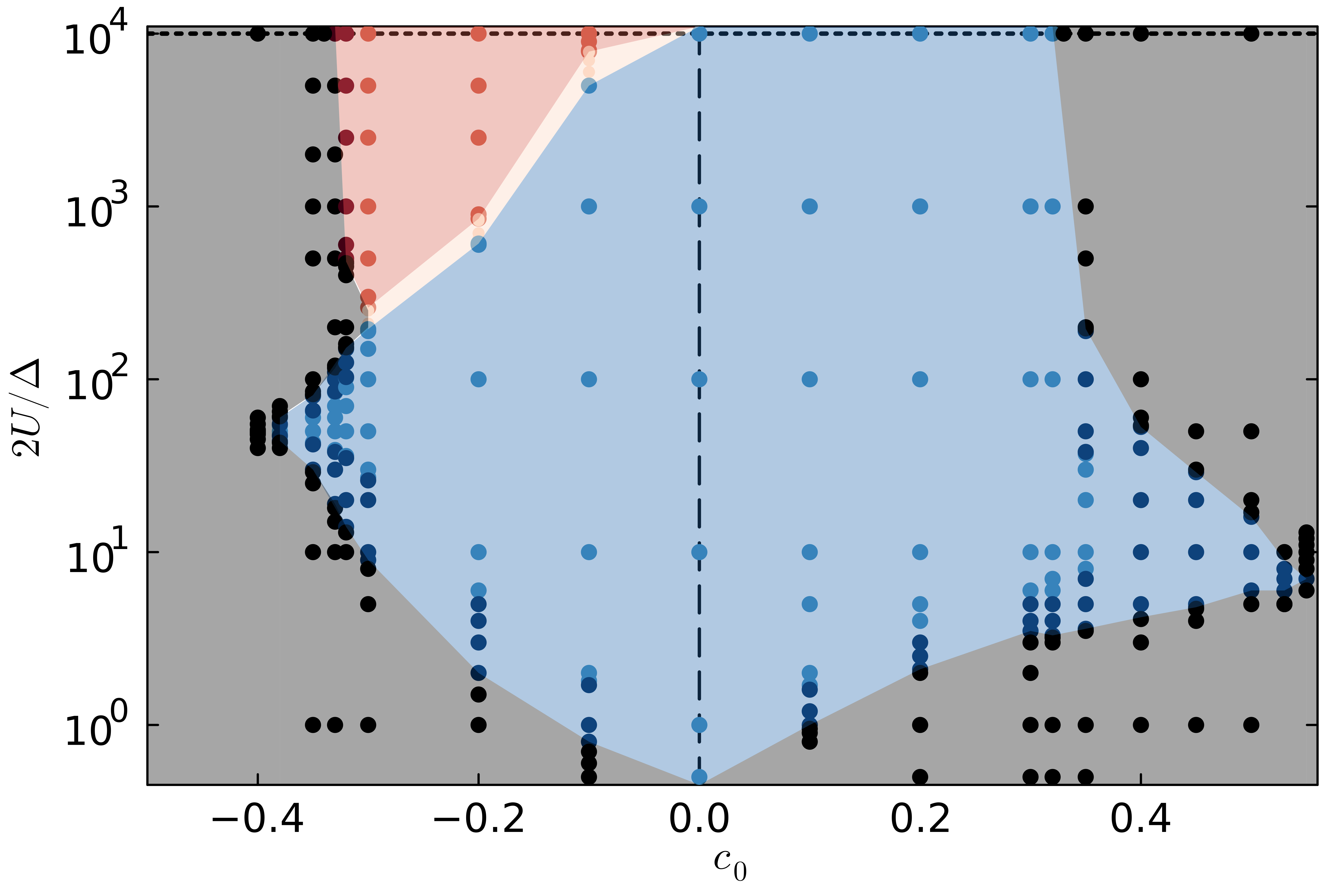}
\caption{Phase diagram of the interacting QBCP flat bands at $\tilde{\alpha}=2.13$ with fixed $m_z=1$ (band gap $\Delta=2m_z$). The symbols are the calculations with $N_x=3,N_y=5$. The red area denotes the FCI with inverted many-body Chern number, which originated mainly from the states of the higher single-particle band, while the blue area represents the ordinary FCI mainly contributed from the states in the lower single-particle band. The white area is the crossover region with mixed FCI states from the two bands. Black area represents the FL phase. Vertical dashed line marks the $c_0=0$ chiral limit while horizontal dotted line marks $U\to \infty$ limit. Dark (light) red and blue mark the FCI states with the spread in the FCI ground state energies larger (smaller) than the many-body gap, indicating more (less) stable FCI states. }
\label{fig:phase}
\end{center}
\end{figure}

In this letter, we present an example where the many-body Chern number or Hall conductance deviates from the expected $\sigma_{xy}=C\nu e^2/h$ due to the quantum metric. We examine a model system featuring two distinct single-particle topological bands with $C=\pm 1$. Under strong Coulomb interaction, electrons preferentially populate the band with a lower quantum metric, despite its higher energy at the single-particle level. Consequently, the many-body Chern number diverges from that predicted by the filling of the lower energy single-particle band. This demonstrates the important role of the quantum metric in FCI phenomena: It not only determines the emergence of the FCI states but also affects the resulting many-body Chern number. Furthermore, the quantum metric generates dispersion through interaction, and causes reentrant transition from FCI to Fermi liquid (FL). The multiple roles of the quantum metric give rise to the rich phase diagram, Fig. \ref{fig:phase}, when tuning the quantum metric through the model parameters and interaction strength.


\textit{Model.} To establish the connection between the quantum geometry and the many-body quantum state, it is desirable to have great tunability of the quantum geometry and the band dispersion. For this purpose, we take topological flat bands emerging from the quadratic band crossing point (QBCP) with periodic strain \cite{PhysRevLett.130.216401} as a model system. The physics unraveled here is more general and is applicable to a broad class of systems including twisted bilayer graphene and twisted MoTe\textsubscript{2}. The QBCP in 2D is described by the continuum Hamiltonian,
\begin{equation}
    H_\Gamma(\mathbf{k}) = -[c_0k^2\sigma_0-(k_x^2-k_y^2)\sigma_x-2k_xk_y\sigma_y+m_z\sigma_z],
\end{equation}
where $\sigma_\alpha$ are the Pauli matrices and $\sigma_0$ is the identity matrix. All quantities are made dimensionless through proper renormalization. As one lattice realization, the QBCP can emerge as a low-energy theory near the $\Gamma$ point in a kagome lattice.
Adding next-nearest-neighbor hopping in the kagome lattice gives control of the parameter $c_0$ while an extra phase in the nearest neighbor hopping gives rise to $m_z$, which breaks time-reversal symmetry \cite{sarkar2023symmetry}. 
In the chiral limit $c_0=0$ and $m_z=0$, $\{H_\Gamma(\mathbf{k}),\sigma_z\}=0$, the spectrum is particle-hole symmetric.
To achieve topological flat bands, we introduce periodic strain to the QBCP $ H(\mathbf{r}) =  H_\Gamma(\mathbf{k})+A_x(\mathbf{r})\sigma_x+A_y(\mathbf{r})\sigma_y$. In the chiral basis, $k_\alpha\to -i \partial_\alpha$, $z\equiv x+iy, \quad \tilde{A}=A_x-iA_y, \quad \partial_z =\frac{1}{2}(\partial_x-i\partial_y)$, we have
\begin{equation}
    \begin{split}
        H(\mathbf{r}) = \begin{pmatrix}
            4c_0\partial_z\partial_{\bar{z}}-m_z & 4\partial_z^2+\tilde{A} \\
            4\partial_{\bar{z}}^2+\tilde{A}^* & 4c_0\partial_z\partial_{\bar{z}}+m_z
        \end{pmatrix}.
    \end{split}
\end{equation}
Within the first-harmonic approximation \cite{PhysRevLett.130.216401}, the strain field has the form $\tilde{A}(\mathbf{r})=\frac{\alpha^2}{2} \sum_{n=1}^3 \omega^{n-1}\cos(\mathbf{G}_n\cdot \mathbf{r}+\phi)$, where $\alpha$ is the strain strength, $\omega=e^{2\pi i/3}$, $\mathbf{G}_1=\frac{4\pi}{\sqrt{3}}(0,1)$,$\mathbf{G}_2=\frac{4\pi}{\sqrt{3}}(-\sqrt{3}/2,-1/2)$, $\mathbf{G}_3=\frac{4\pi}{\sqrt{3}}(\sqrt{3}/2,-1/2)$ are reciprocal lattice vectors.  In the chiral limit $c_0=0$, topological exact flat bands appear at $E=0$ with Chern number $C=\pm 1$ and ideal quantum metric at specific strain strength $\tilde{\alpha}=\alpha/|\mathbf{G}^m|=0.79,\ 2.13,\ 3.52,\ldots$. The Chern bands are sublattice polarized, which is defined as the eigenstate of $\sigma_z$. The introduction of a finite  $m_z$ breaks time-reversal symmetry, raising the energy of one Chern band while lowering the other, with the sign of  $m_z$  dictating the sign of the Hall conductance in the noninteracting regime. However, in the regime of strong interaction, especially for the FCI, the quantum metric can reverse the sign of the Hall conductance from that expected in the noninteracting scenario.
Beyond the chiral limit, we can adjust the band dispersion and quantum metric of the two low-energy topological flat bands by $c_0$, deforming the Berry curvature and violating the trace condition \cite{PhysRevB.90.165139}(see SI Sec. I).

\begin{figure}[b!]
\begin{center}
\setlabel{pos=nw,fontsize=\large,labelbox=false}
\includegraphics[width = 0.49\textwidth]{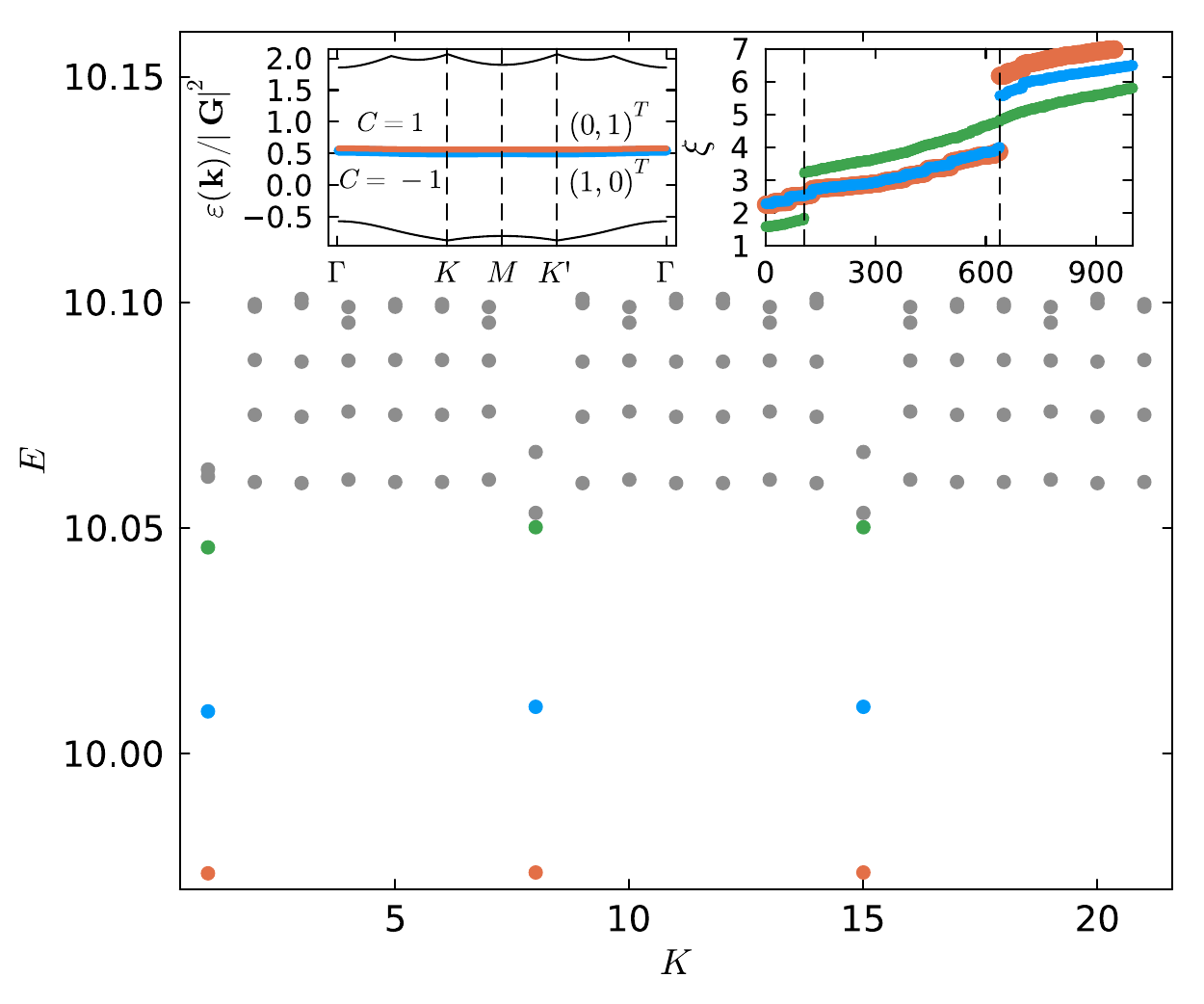}
\caption{Many-body state spectrum of the inverted FCI at $c_0=-0.25,m_z=1$ with grid size $N=21$ (SI Sec. II) and $Q=7$ electrons, with infinite $U$ (red region in Fig. \ref{fig:phase}). Left inset: The single-particle dispersion with $C=-1$ (blue) band (A sublattice ploarized $(1,0)^T$)  lower than $C=1$ (red) band (B sublattice ploarized $(0,1)^T$). Right inset: The particle entanglement spectrum $\xi$ of the lowest $3$ states with a many-body $C_{\mathrm{MB}}=1/3$ and the next lowest $3$ states with many-body $C_\mathrm{MB}=-1/3$ (see SI Sec. V for the many-body Chern number calculation). Vertical dashed line marks the analytical spectrum gap for the CDW ($N_\xi = 105$) and the FCI ($N_\xi=637$) (SI Sec. VI). 
}
\label{fig:spectrum}
\end{center}
\end{figure}

To study the interaction effect, we project the density operator onto the two flat bands (red and blue bands in the inset of Fig. \ref{fig:spectrum}) and carry out an exact diagonalization (ED) calculation. The interacting Hamiltonian becomes \cite{lee2019theory,PhysRevLett.125.226401,PhysRevResearch.3.L032070}
\begin{equation}
    \begin{split}
        H_\mathrm{int} = \sum_{\mathbf{k},\tau} (\epsilon_{\mathbf{k},\tau}-\mu) c^\dagger_{\mathbf{k},\tau} c_{\mathbf{k},\tau}+\frac{1}{2A} \sum_{\mathbf{q}} \rho(\mathbf{q})V(\mathbf{q})\rho(-\mathbf{q}),
    \end{split}
\end{equation}
where $A$ is the volume (area) of the Brillouin zone (BZ) and 
$V(\mathbf{q})=4\pi U \tanh(qd)/(\sqrt{3}qa)$ 
is the screened Coulomb potential, where $d=2a$ is the separation between the electrode and the 2D system, and $a$ is the period of the strain \cite{PhysRevResearch.3.L032070}. We label the bands by their Chern numbers $\tau=\pm 1$. The projected density operator is 
\begin{equation}
\begin{split}
    \rho(\mathbf{q})&=\sum_{\mathbf{k,k'},\tau,\tau'} \langle \psi_{\mathbf{k},\tau}|e^{i\mathbf{q\cdot r}}|\psi_{\mathbf{k'},\tau'}\rangle c^\dagger_{\mathbf{k},\tau} c_{\mathbf{k+q},\tau'}\\
    &=\sum_{\mathbf{k},\tau,\tau'} \lambda_{\tau,\tau',\mathbf{q}}(\mathbf{k}) c^\dagger_{\mathbf{k},\tau} c_{\mathbf{k+q},\tau'}
\end{split}
\end{equation}
where $\psi_{\mathbf{k},\tau}(\mathbf{r})=e^{i\mathbf{k\cdot r}}u_{\tau,\mathbf{k}}(\mathbf{r})=\frac{1}{\sqrt{\Omega}} \sum_\mathbf{G} e^{i\mathbf{(G+k)\cdot r}}u_{\tau,\mathbf{k}}(\mathbf{G})$ is the Bloch state and $\lambda_{\tau,\tau',\mathbf{q}}(\mathbf{k})=\langle u_{\tau,\mathbf{k}}|u_{\tau',\mathbf{k+q}}\rangle$ is the form factor. We consider the strong interaction limit where $U\gg W_\tau$, $\Delta$, where $W_\tau$ and $\Delta$ are the bandwidth and band gap between the $\tau=\pm 1$ bands. When the interband hybridization is neglected, i.e. the single-particle state is sublattice-polarized,
$\lambda_{\tau,\tau',\mathbf{q}}=\delta_{\tau,\tau'}\lambda_{\tau,\tau,\mathbf{q}}$, the band occupation of electrons is a good quantum number since $[H_\mathrm{int},\tau_z]=0$ where $S^z\equiv \langle \sum_{k,\tau} \tau_z(k,\tau)/2\rangle$ is the band occupation.

\begin{figure}[b!]
\begin{center}
\includegraphics[width = 0.49\textwidth]{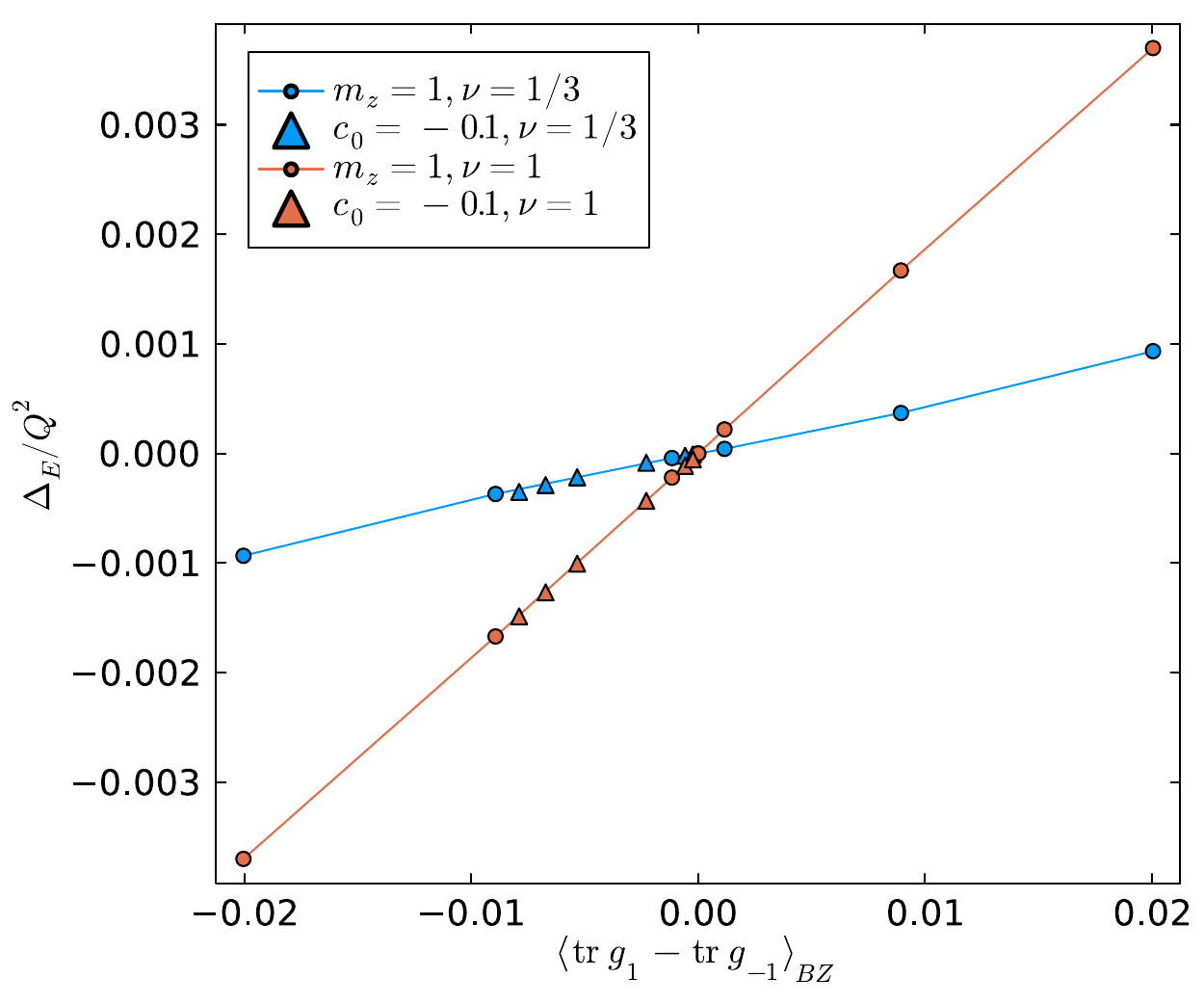}
\caption{Energy difference $\Delta_E\equiv \langle E_{\tau}-E_{-\tau}\rangle_\nu/Q^2$
vs total trace of the quantum metric obtained by ED. $E_\tau$ is the lowest energy of the states associated with the occupation of the band $\tau$, $\langle \cdots \rangle_\nu$ denotes average over nearly degenerate states, $Q$ is the total number of electrons at each filling $\nu$ with a fixed system size $N=21$. Circles represent fixed $m_z$ but with varying $c_0$ while triangles represent fixed $c_0$ but with varying $m_z$. Blue curve is for the $\nu=1/3$ FCI states (averaged over $3$ nearly degenerate states) while red curve is for the $\nu=1$ fully filled Chern insulators.
}
\label{fig:totalEs}
\end{center}
\end{figure}

In the chiral limit $c_0=0$ at the electron filling of $\nu=1/m$ with $m$ an odd integer, the ground state is FCI due to its ideal trace condition. Here we focus on $m=3$, and the conclusion is valid for other values of $m$. The phase diagram obtained by ED is shown in Fig. \ref{fig:phase}, where $c_0$ tunes the quantum metric and the dispersion of the bands. There are two FCI phases (red and blue regions) and the FL phase (gray region). There are two salient features in the phase diagram. 1) In the red regime, the many-body Chern number of the FCI phase is inverted compared with the filling of the lower single-particle band, while in the blue regime, the many-body Chern number aligns with that of the lower single-particle band. 2) When the interaction $U$ increases, there exists a reentrant transition to the FL from the FCI phase around $c_0=\pm 0.4$. In the following, we show that these two features originated from the quantum metric of the single-particle wave function.

In the chiral limit $c_0=0$, there are $6$ degenerate ground states in the ED spectrum, which can be grouped into $2$ sets of FCI states belonging to different band/sublattice polarization. In reality, the system spontaneously selects one band polarization due to symmetry breaking, similar to spontaneous valley polarization in twisted MoTe\textsubscript{2} 
\cite{cai2023signatures,PhysRevResearch.3.L032070} and pentalayer graphene \cite{han2023orbital}. The FCI nature of the many-body states is confirmed by their total momentum $K=1, 8, 15$ according to the Haldane rule \cite{PhysRevLett.67.937,PhysRevB.49.2947}(see SI Sec. II), and their fractional nature can be confirmed by their many-body Chern number and particle entanglement spectrum gap \cite{PhysRevX.1.021014}. From the density distribution $\langle n_{\mathbf{k},\tau}\rangle=\langle c^\dagger_{\mathbf{k},\tau}c_{\mathbf{k},\tau}\rangle$, the lowest $6$ states have a uniform density $1/3$ in each band. 

When chiral symmetry is broken ($c_0\ne 0$), the band occupation is no longer a good quantum number. We consider the weak chiral-symmetry breaking region where the single-particle wave function is still predominantly sublattice polarized and the interband hybridization is small with the help of a sublattice polarization field, $m_z=1$. In the red region in Fig. \ref{fig:phase}, the single-particle band with $C=-1$ lies below the band with $C=1$. Surprisingly, the ground states are the $3$ degenerate FCI states with many-body Chern numbers equal to $C_\mathrm{MB}=1/3$, which is not expected from the single-particle band. The lowest FCI states have band occupation $S^z\approx Q/2$, which implies that doped electrons occupy the band with $C=1$, which has a higher single-particle band energy. In the many-body spectrum, the next $3$ lowest degenerate states (blue in Fig. \ref{fig:spectrum}) are also FCI states with many-body Chern number $C_\mathrm{MB}=-1/3$ and have band occupation $S^z\approx-Q/2$. These FCI states correspond to the partial filling of the $C=-1$ lower single particle band. Interestingly, the third set of $3$ states (green circles in Fig. \ref{fig:spectrum}) exhibit a particle entanglement spectrum gap similar to that of charge density wave (CDW) states, and the density-density correlation peaks in momentum space \cite{li2024contrasting,PhysRevLett.127.246403}.
 However, these states never become ground states in the QBCP flat bands when tuning $c_0$.

Our ED results show that, for a fixed $m_z$, the many-body Chern number and the sign of the Hall conductance may flip compared to the single-particle band Chern number as the system moves away from the ideal quantum geometry $c_0=0$. This inversion of many-body Chern number also occurs for other fractional fillings, i.e., $\nu=2/3,2/5,2/7$ (SI Sec. IV). To understand this inversion effect, we employ the Hartree-Fock (HF) mean field approximation at integer filling $\nu=1$ without interband hybridization. The role of the interband hybridization is discussed in SI Sec. VIII. The self-consistent Hartree-Fock approximation gives
\begin{equation}
    \begin{split}
        E_{\mathbf{k},\tau} &= \epsilon_{\mathbf{k},\tau}-\mu+\Delta_H(\mathbf{k},\tau)+\Delta_F(\mathbf{k},\tau)
    \end{split}
\end{equation}
where $\Delta_H(\mathbf{k},\tau),\Delta_F(\mathbf{k},\tau)$ are the Hartree and Fock energies 
\begin{equation}
    \begin{split}
        \Delta_H(\mathbf{k},\tau)&=\frac{1}{A}\sum_{\mathbf{q}=n\mathbf{G}} \lambda_{\tau,\mathbf{q}}(\mathbf{k})V(\mathbf{q}) \sum_{\mathbf{k'}} \lambda_{\tau',-\mathbf{q}}(\mathbf{k'})\\
        \Delta_F(\mathbf{k},\tau)&=-\frac{1}{A}\sum_{\mathbf{q,k'=k+q}} \lambda_{\tau,\mathbf{q}}(\mathbf{k})V(\mathbf{q})\lambda_{\tau,-\mathbf{q}}(\mathbf{k'}).
    \end{split}
\end{equation}
For a fast decay $V(q)$, we approximate $V(n\mathbf{G})=V\delta_{n0}$, and $\Delta_H(\mathbf{k},\tau)\approx QV/A$. The Hartree term depends only on density and is independent of the quantum metric of each band.

On the other hand, the Fock contribution can be written as
\begin{equation}
\begin{split}
    \Delta_F(\mathbf{k},\tau)
    &\approx -\frac{V}{A}\sum_{\mathbf{q},q<q_c} |\lambda_{\tau,\mathbf{q}}(\mathbf{k})|^2f(E_{\mathbf{k},\tau}),
\end{split}
\end{equation}
where we have used $\lambda_{\tau,-\mathbf{q}}(\mathbf{k+q})=\lambda_{\tau,\mathbf{q}}^*(\mathbf{k})$ and $\tau=\tau'$. It becomes clear that the Fock energy is connected to the Hilbert-Schmidt quantum distance of each band, $s_\tau^2(\mathbf{k,k+q})\equiv 1-|\langle u_{\tau,\mathbf{k}}|u_{\tau,\mathbf{k+q}}\rangle |^2=1-|\lambda_{\tau,\mathbf{q}}(\mathbf{k})|^2$. For a fast decay $V(q)$, we can expand $s_\tau$ for small $q$, $s_\tau^2(\mathbf{k,k+q})\approx g_{\tau,\mu\nu} q_\mu q_\nu$.
Then the Fock energy becomes
\begin{equation}\label{eq:fock}
    \begin{split}
        \Delta_F(\mathbf{k},\tau) 
        &\approx \mathcal{O}(1) +\frac{\pi V_\mathrm{int}}{A}f(E_{\mathbf{k},\tau})\mathrm{tr}[g_\tau(\mathbf{k})],
    \end{split}
\end{equation}
with some constant $\mathcal{O}(1)$ and $V_{\mathrm{int}}\equiv \int \mathrm{d}q q^3 V(q)$ (see SI Sec. VII). Thus, the Fock energy depends on the quantum metric of the bands.
In the chiral limit, the trace condition is satisfied $\mathrm{tr} [g(\mathbf{k},\tau)] = |F_{x,y}(\mathbf{k},\tau)|$, 
and $\int d\mathbf{k}|F_{xy}(\mathbf{k},\tau)|=1$. The total Fock energy is the same for the $\tau=\pm 1$ band. However, when the chiral symmetry is broken, $\Delta_F(\mathbf{k},\tau) \propto \mathrm{tr}[g_\tau(\mathbf{k})]$ splits the $\tau=\pm$ bands depending on their quantum metric. The band with a smaller total trace of the quantum metric becomes lower in energy 
in the strong-interacting limit. Note that $\mathrm{tr}[g_\tau(\mathbf{k})]$ is bounded below by the Berry curvature of the band, so it cannot be zero for a topological flat band. 

\begin{figure}[t!]
\begin{center}
\includegraphics[width = 0.45\textwidth]{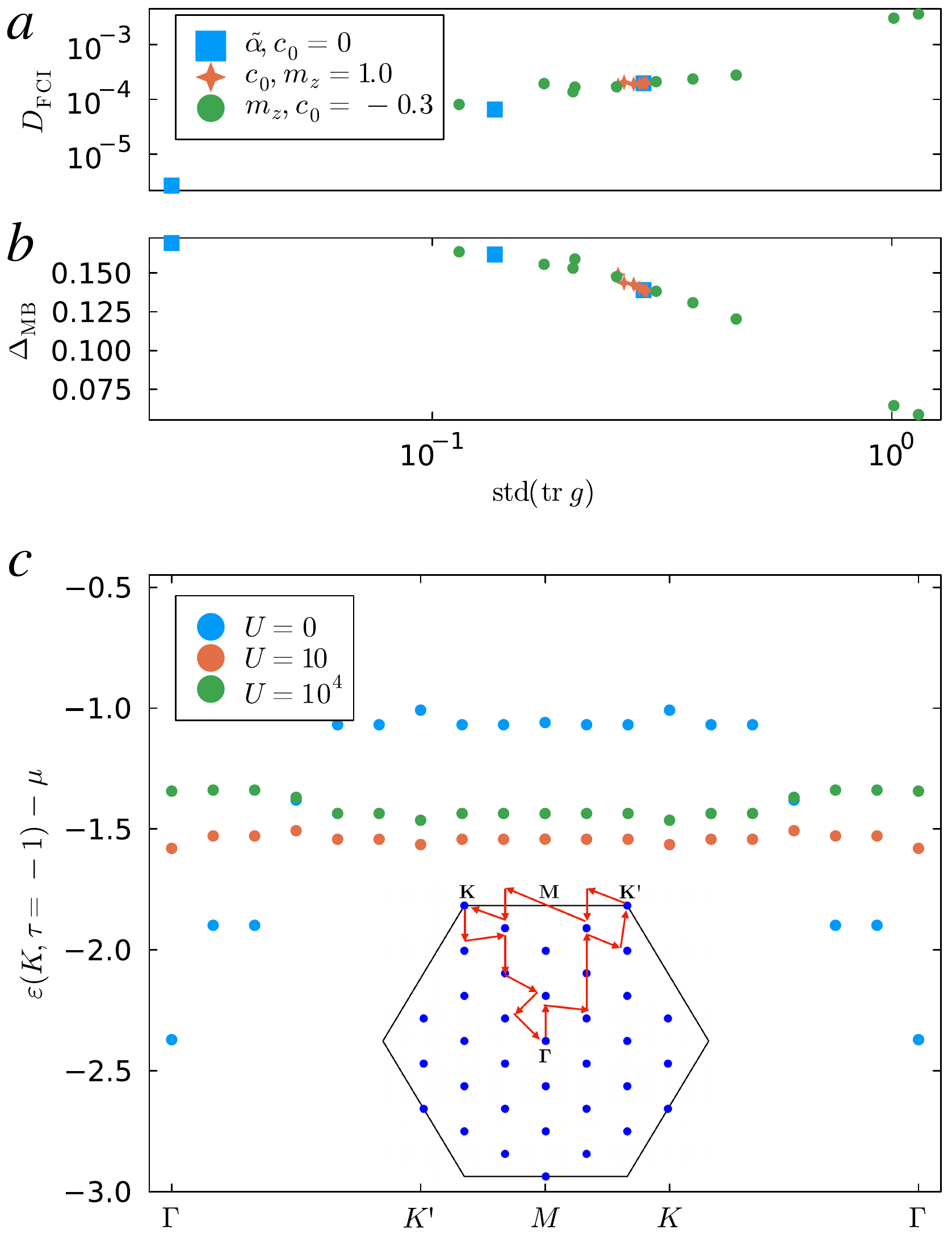}
\caption{Quality of FCI states versus standard deviation of $\mathrm{tr}[g]$ obtained by the ED without interband hybridization. Blue squares are from $3$ different magic parameters $\tilde{\alpha}$ in the chiral limit $c_0=0$. Red and green data are obtained at the first magic parameter $\tilde{\alpha}_1=0.79$. (a) Spread of the ground state energies for the 3 quasi-degenerate FCI states $D_\mathrm{FCI}\equiv E_3-E_1$. (b) Many-body gap of the FCI states $\Delta_\mathrm{MB}\equiv E_4-E_1$, where $E_n$ denotes $n$-th many-body energy. (c) The lower HF band dispersion at the cut $c_0=0.35$ in Fig. \ref{fig:phase} where the reentrant transition to FL occurs. The dispersion is offset by the average band energy $\mu\equiv \langle \epsilon \rangle$, and finite $U$ results are renormalized by $U$. Here the system size is $N_x=N_y=6$.}
\label{fig:FCIquality}
\end{center}
\end{figure}

We verify the above analysis numerically both for $\nu=1$ and fractional fillings. Indeed, the energy of the occupied band is proportional to the trace of its quantum metric, and the energy difference  $\Delta_{E}\equiv E_{C=1}-E_{C=-1}\propto \sum_{\mathbf{k}}\mathrm{tr}[g(\mathbf{k},1)-g(\mathbf{k},-1)]$, as shown in Fig. \ref{fig:totalEs}. One may argue that the HF analysis for a full filling of the band at $\nu=1$ (Eq. \ref{eq:fock}) also applies to FCI because electrons in FCI occupy all momentum points equally. When $c_0 <0$ and $m_z >0$, the single particle band with a higher energy carries a smaller total trace of the quantum metric. As a consequence, the interaction inverts the band energy when it is occupied, which results in an inverted many-body Chern number compared with the expectation of a partial filling of the lower single-particle band.

The first role of the quantum metric in determining the stability of FCI is through the trace condition, i.e., the FCI is favored when $\mathrm{tr}[g(k)]=F_{xy}(k)$. The second role of the quantum metric is to modify the single-particle band alignment and cause the inversion of the many-body Chern number compared with the single-particle band topology. Since the contribution of the quantum metric to the interaction energy is momentum dependent, it also modifies the band dispersion, $\Delta_F(\mathbf{k},\tau)\sim \mathrm{tr}[g_\tau(\mathbf{k})]$, which affects the stability of the FCI. The variation of the quantum metric in the momentum space induces dispersion in the band structure. Here, we introduce the standard deviation of the quantum metric $\mathrm{std}(\mathrm{tr}\; g)\equiv \sqrt{\langle (\mathrm{tr}\; g)^2\rangle-\langle \mathrm{tr}\; g\rangle^2}$ and study its role in the stability of FCI. We define two quantities to quantify the quality of the FCI states: (a) $D_\mathrm{FCI}$ is the spread of the otherwise degenerate FCI energy states in the ideal limit; (b) $\Delta_\mathrm{MB}$ is the many-body gap of the FCI state. The smaller the $D_\mathrm{FCI}$ and the larger the $\Delta_\mathrm{MB}$, the more robust the FCI. In Fig. \ref{fig:FCIquality}, the quality of the FCI states is shown to be inversely correlated with the standard deviation of the quantum metric, regardless of the source of deviation ($\tilde{\alpha},c_0$ or $m_z$). This is consistent with the expectation that a more dispersive band caused by a large variation of $\mathrm{tr}[g(\mathbf{k})]$ disfavors FCI. As the variation of $\mathrm{tr}[g(\mathbf{k})]$ increases further, the Fock correction to the bands is expected to destroy the FCI states and stabilize the FL \cite{PhysRevResearch.5.L012015}.

The quantum metric-induced band dispersion also naturally explains the reentrant Fermi liquid state in the phase diagram, Fig. \ref{fig:phase}. Around $c_0=\pm 0.4$, upon increasing $U$, the system first transits from FL to FCI and then FL. The first transition into FCI is natural since FCI requires interaction dominance over the kinetic energy of electrons. Our calculations show that the quantum metric also helps flatten the dispersion at intermediate $U$, as shown in Fig. \ref{fig:FCIquality}(c). The single particle band has a strong dispersion around the $\Gamma$ point, which is compensated by the Fock energy at an intermediate $U$. For a large $U$, the HF band becomes dispersive again due to the quantum metric, which induces the second transition from FCI to FL.

In summary, we explore the stability of FCI beyond ideal quantum geometries, particularly for flatbands in systems having QBCP under periodic strain. Using ED and Hartree-Fock calculations, we reveal the significant roles of the quantum metric in stabilizing the FCI. We find that the quantum metric causes the many-body Chern number of the FCI to deviate significantly from the expected value when partially filling the lower energy single-particle band. Additionally, the variation of the quantum metric in momentum space induces band dispersion through interaction, which impacts the stability of the FCI. As one manifestation, we show the reentrance of FL from the FCI state as the interaction strength increases. Our results have broad implications. For instance, in single-particle band configurations with $C=0$ and $C=1$, it is possible to have an interaction-induced transition from CDW to FCI if electrons prefer to fill the $C=0$ or $C=1$ bands depending on the quantum metric. In this letter, we highlight the role of quantum metric in stabilizing the FCI in the strongly correlated topological flat bands and showcase how the resultant FCI deviates from the expectation of partially filling a single-particle band as also exemplified in pentalayer graphene.  

\textit{Acknowledgments.} We thank Heqiu Li, Xi Dai and Emil Bergholtz for useful discussions. The work at LANL is partially supported by the U.S. Department of Energy (DOE) National Nuclear Security Administration (NNSA) under Contract No. 89233218CNA000001 through the Laboratory Directed Research and Development (LDRD) Program and was performed, in part, at the Center for Integrated Nanotechnologies, an Office of Science User Facility operated for the DOE Office of Science, under user Proposals No. 2018BU0010 and No. 2018BU0083. The work at University of Michigan is partially supported by the Air Force Office of Scientific Research through the Multidisciplinary University Research Initiative, Award No. FA9550-23-1-0334, and the Office of Naval Research Multidisciplinary University Research Initiatives (MURI) No. N00014- 20-1-2479, and Award No. N00014-21-1-2770, and by the Gordon and Betty Moore Foundation Award No. N031710.

\bibliography{refs}

\begin{thebibliography}{64}%
\makeatletter
\providecommand \@ifxundefined [1]{%
 \@ifx{#1\undefined}
}%
\providecommand \@ifnum [1]{%
 \ifnum #1\expandafter \@firstoftwo
 \else \expandafter \@secondoftwo
 \fi
}%
\providecommand \@ifx [1]{%
 \ifx #1\expandafter \@firstoftwo
 \else \expandafter \@secondoftwo
 \fi
}%
\providecommand \natexlab [1]{#1}%
\providecommand \enquote  [1]{``#1''}%
\providecommand \bibnamefont  [1]{#1}%
\providecommand \bibfnamefont [1]{#1}%
\providecommand \citenamefont [1]{#1}%
\providecommand \href@noop [0]{\@secondoftwo}%
\providecommand \href [0]{\begingroup \@sanitize@url \@href}%
\providecommand \@href[1]{\@@startlink{#1}\@@href}%
\providecommand \@@href[1]{\endgroup#1\@@endlink}%
\providecommand \@sanitize@url [0]{\catcode `\\12\catcode `\$12\catcode `\&12\catcode `\#12\catcode `\^12\catcode `\_12\catcode `\%12\relax}%
\providecommand \@@startlink[1]{}%
\providecommand \@@endlink[0]{}%
\providecommand \url  [0]{\begingroup\@sanitize@url \@url }%
\providecommand \@url [1]{\endgroup\@href {#1}{\urlprefix }}%
\providecommand \urlprefix  [0]{URL }%
\providecommand \Eprint [0]{\href }%
\providecommand \doibase [0]{https://doi.org/}%
\providecommand \selectlanguage [0]{\@gobble}%
\providecommand \bibinfo  [0]{\@secondoftwo}%
\providecommand \bibfield  [0]{\@secondoftwo}%
\providecommand \translation [1]{[#1]}%
\providecommand \BibitemOpen [0]{}%
\providecommand \bibitemStop [0]{}%
\providecommand \bibitemNoStop [0]{.\EOS\space}%
\providecommand \EOS [0]{\spacefactor3000\relax}%
\providecommand \BibitemShut  [1]{\csname bibitem#1\endcsname}%
\let\auto@bib@innerbib\@empty
\bibitem [{\citenamefont {Tsui}\ \emph {et~al.}(1982)\citenamefont {Tsui}, \citenamefont {Stormer},\ and\ \citenamefont {Gossard}}]{PhysRevLett.48.1559}%
  \BibitemOpen
  \bibfield  {author} {\bibinfo {author} {\bibfnamefont {D.~C.}\ \bibnamefont {Tsui}}, \bibinfo {author} {\bibfnamefont {H.~L.}\ \bibnamefont {Stormer}},\ and\ \bibinfo {author} {\bibfnamefont {A.~C.}\ \bibnamefont {Gossard}},\ }\bibfield  {title} {\bibinfo {title} {Two-dimensional magnetotransport in the extreme quantum limit},\ }\href {https://doi.org/10.1103/PhysRevLett.48.1559} {\bibfield  {journal} {\bibinfo  {journal} {Phys. Rev. Lett.}\ }\textbf {\bibinfo {volume} {48}},\ \bibinfo {pages} {1559} (\bibinfo {year} {1982})}\BibitemShut {NoStop}%
\bibitem [{\citenamefont {Stormer}(1999)}]{RevModPhys.71.875}%
  \BibitemOpen
  \bibfield  {author} {\bibinfo {author} {\bibfnamefont {H.~L.}\ \bibnamefont {Stormer}},\ }\bibfield  {title} {\bibinfo {title} {Nobel lecture: The fractional quantum hall effect},\ }\href {https://doi.org/10.1103/RevModPhys.71.875} {\bibfield  {journal} {\bibinfo  {journal} {Rev. Mod. Phys.}\ }\textbf {\bibinfo {volume} {71}},\ \bibinfo {pages} {875} (\bibinfo {year} {1999})}\BibitemShut {NoStop}%
\bibitem [{\citenamefont {Laughlin}(1983)}]{PhysRevLett.50.1395}%
  \BibitemOpen
  \bibfield  {author} {\bibinfo {author} {\bibfnamefont {R.~B.}\ \bibnamefont {Laughlin}},\ }\bibfield  {title} {\bibinfo {title} {Anomalous quantum hall effect: An incompressible quantum fluid with fractionally charged excitations},\ }\href {https://doi.org/10.1103/PhysRevLett.50.1395} {\bibfield  {journal} {\bibinfo  {journal} {Phys. Rev. Lett.}\ }\textbf {\bibinfo {volume} {50}},\ \bibinfo {pages} {1395} (\bibinfo {year} {1983})}\BibitemShut {NoStop}%
\bibitem [{\citenamefont {Pan}\ \emph {et~al.}(2003)\citenamefont {Pan}, \citenamefont {Stormer}, \citenamefont {Tsui}, \citenamefont {Pfeiffer}, \citenamefont {Baldwin},\ and\ \citenamefont {West}}]{PhysRevLett.90.016801}%
  \BibitemOpen
  \bibfield  {author} {\bibinfo {author} {\bibfnamefont {W.}~\bibnamefont {Pan}}, \bibinfo {author} {\bibfnamefont {H.~L.}\ \bibnamefont {Stormer}}, \bibinfo {author} {\bibfnamefont {D.~C.}\ \bibnamefont {Tsui}}, \bibinfo {author} {\bibfnamefont {L.~N.}\ \bibnamefont {Pfeiffer}}, \bibinfo {author} {\bibfnamefont {K.~W.}\ \bibnamefont {Baldwin}},\ and\ \bibinfo {author} {\bibfnamefont {K.~W.}\ \bibnamefont {West}},\ }\bibfield  {title} {\bibinfo {title} {Fractional quantum hall effect of composite fermions},\ }\href {https://doi.org/10.1103/PhysRevLett.90.016801} {\bibfield  {journal} {\bibinfo  {journal} {Phys. Rev. Lett.}\ }\textbf {\bibinfo {volume} {90}},\ \bibinfo {pages} {016801} (\bibinfo {year} {2003})}\BibitemShut {NoStop}%
\bibitem [{\citenamefont {TRIVEDI}\ and\ \citenamefont {JAIN}(1991)}]{trivedi1991numerical}%
  \BibitemOpen
  \bibfield  {author} {\bibinfo {author} {\bibfnamefont {N.}~\bibnamefont {TRIVEDI}}\ and\ \bibinfo {author} {\bibfnamefont {J.}~\bibnamefont {JAIN}},\ }\bibfield  {title} {\bibinfo {title} {Numerical study of jastrow-slater trial states for the fractional quantum hall effect},\ }\href {https://doi.org/10.1142/S0217984991000599} {\bibfield  {journal} {\bibinfo  {journal} {Modern Physics Letters B}\ }\textbf {\bibinfo {volume} {05}},\ \bibinfo {pages} {503} (\bibinfo {year} {1991})}\BibitemShut {NoStop}%
\bibitem [{\citenamefont {Jain}(1989)}]{PhysRevLett.63.199}%
  \BibitemOpen
  \bibfield  {author} {\bibinfo {author} {\bibfnamefont {J.~K.}\ \bibnamefont {Jain}},\ }\bibfield  {title} {\bibinfo {title} {Composite-fermion approach for the fractional quantum hall effect},\ }\href {https://doi.org/10.1103/PhysRevLett.63.199} {\bibfield  {journal} {\bibinfo  {journal} {Phys. Rev. Lett.}\ }\textbf {\bibinfo {volume} {63}},\ \bibinfo {pages} {199} (\bibinfo {year} {1989})}\BibitemShut {NoStop}%
\bibitem [{\citenamefont {Jain}(2007)}]{jain2007composite}%
  \BibitemOpen
  \bibfield  {author} {\bibinfo {author} {\bibfnamefont {J.~K.}\ \bibnamefont {Jain}},\ }\href@noop {} {\emph {\bibinfo {title} {Composite fermions}}}\ (\bibinfo  {publisher} {Cambridge University Press},\ \bibinfo {year} {2007})\BibitemShut {NoStop}%
\bibitem [{\citenamefont {Fremling}\ \emph {et~al.}(2018)\citenamefont {Fremling}, \citenamefont {Moran}, \citenamefont {Slingerland},\ and\ \citenamefont {Simon}}]{PhysRevB.97.035149}%
  \BibitemOpen
  \bibfield  {author} {\bibinfo {author} {\bibfnamefont {M.}~\bibnamefont {Fremling}}, \bibinfo {author} {\bibfnamefont {N.}~\bibnamefont {Moran}}, \bibinfo {author} {\bibfnamefont {J.~K.}\ \bibnamefont {Slingerland}},\ and\ \bibinfo {author} {\bibfnamefont {S.~H.}\ \bibnamefont {Simon}},\ }\bibfield  {title} {\bibinfo {title} {Trial wave functions for a composite fermi liquid on a torus},\ }\href {https://doi.org/10.1103/PhysRevB.97.035149} {\bibfield  {journal} {\bibinfo  {journal} {Phys. Rev. B}\ }\textbf {\bibinfo {volume} {97}},\ \bibinfo {pages} {035149} (\bibinfo {year} {2018})}\BibitemShut {NoStop}%
\bibitem [{\citenamefont {Park}\ \emph {et~al.}(2023)\citenamefont {Park}, \citenamefont {Cai}, \citenamefont {Anderson}, \citenamefont {Zhang}, \citenamefont {Zhu}, \citenamefont {Liu}, \citenamefont {Wang}, \citenamefont {Holtzmann}, \citenamefont {Hu}, \citenamefont {Liu} \emph {et~al.}}]{park2023observation}%
  \BibitemOpen
  \bibfield  {author} {\bibinfo {author} {\bibfnamefont {H.}~\bibnamefont {Park}}, \bibinfo {author} {\bibfnamefont {J.}~\bibnamefont {Cai}}, \bibinfo {author} {\bibfnamefont {E.}~\bibnamefont {Anderson}}, \bibinfo {author} {\bibfnamefont {Y.}~\bibnamefont {Zhang}}, \bibinfo {author} {\bibfnamefont {J.}~\bibnamefont {Zhu}}, \bibinfo {author} {\bibfnamefont {X.}~\bibnamefont {Liu}}, \bibinfo {author} {\bibfnamefont {C.}~\bibnamefont {Wang}}, \bibinfo {author} {\bibfnamefont {W.}~\bibnamefont {Holtzmann}}, \bibinfo {author} {\bibfnamefont {C.}~\bibnamefont {Hu}}, \bibinfo {author} {\bibfnamefont {Z.}~\bibnamefont {Liu}}, \emph {et~al.},\ }\bibfield  {title} {\bibinfo {title} {Observation of fractionally quantized anomalous hall effect},\ }\href {https://www.nature.com/articles/s41586-023-06536-0} {\bibfield  {journal} {\bibinfo  {journal} {Nature}\ }\textbf {\bibinfo {volume} {622}},\ \bibinfo {pages} {74} (\bibinfo {year} {2023})}\BibitemShut {NoStop}%
\bibitem [{\citenamefont {Cai}\ \emph {et~al.}(2023)\citenamefont {Cai}, \citenamefont {Anderson}, \citenamefont {Wang}, \citenamefont {Zhang}, \citenamefont {Liu}, \citenamefont {Holtzmann}, \citenamefont {Zhang}, \citenamefont {Fan}, \citenamefont {Taniguchi}, \citenamefont {Watanabe}, \citenamefont {Ran}, \citenamefont {Cao}, \citenamefont {Fu}, \citenamefont {Xiao}, \citenamefont {Yao},\ and\ \citenamefont {Xu}}]{cai2023signatures}%
  \BibitemOpen
  \bibfield  {author} {\bibinfo {author} {\bibfnamefont {J.}~\bibnamefont {Cai}}, \bibinfo {author} {\bibfnamefont {E.}~\bibnamefont {Anderson}}, \bibinfo {author} {\bibfnamefont {C.}~\bibnamefont {Wang}}, \bibinfo {author} {\bibfnamefont {X.}~\bibnamefont {Zhang}}, \bibinfo {author} {\bibfnamefont {X.}~\bibnamefont {Liu}}, \bibinfo {author} {\bibfnamefont {W.}~\bibnamefont {Holtzmann}}, \bibinfo {author} {\bibfnamefont {Y.}~\bibnamefont {Zhang}}, \bibinfo {author} {\bibfnamefont {F.}~\bibnamefont {Fan}}, \bibinfo {author} {\bibfnamefont {T.}~\bibnamefont {Taniguchi}}, \bibinfo {author} {\bibfnamefont {K.}~\bibnamefont {Watanabe}}, \bibinfo {author} {\bibfnamefont {Y.}~\bibnamefont {Ran}}, \bibinfo {author} {\bibfnamefont {T.}~\bibnamefont {Cao}}, \bibinfo {author} {\bibfnamefont {L.}~\bibnamefont {Fu}}, \bibinfo {author} {\bibfnamefont {D.}~\bibnamefont {Xiao}}, \bibinfo {author} {\bibfnamefont {W.}~\bibnamefont {Yao}},\ and\ \bibinfo {author} {\bibfnamefont {X.}~\bibnamefont {Xu}},\ }\bibfield  {title}
  {\bibinfo {title} {Signatures of fractional quantum anomalous hall states in twisted {MoTe2}},\ }\href {http://dx.doi.org/10.1038/s41586-023-06289-w} {\bibfield  {journal} {\bibinfo  {journal} {Nature}\ }\textbf {\bibinfo {volume} {622}},\ \bibinfo {pages} {63} (\bibinfo {year} {2023})}\BibitemShut {NoStop}%
\bibitem [{\citenamefont {Zeng}\ \emph {et~al.}(2023)\citenamefont {Zeng}, \citenamefont {Xia}, \citenamefont {Kang}, \citenamefont {Zhu}, \citenamefont {Kn{\"u}ppel}, \citenamefont {Vaswani}, \citenamefont {Watanabe}, \citenamefont {Taniguchi}, \citenamefont {Mak},\ and\ \citenamefont {Shan}}]{zeng2023thermodynamic}%
  \BibitemOpen
  \bibfield  {author} {\bibinfo {author} {\bibfnamefont {Y.}~\bibnamefont {Zeng}}, \bibinfo {author} {\bibfnamefont {Z.}~\bibnamefont {Xia}}, \bibinfo {author} {\bibfnamefont {K.}~\bibnamefont {Kang}}, \bibinfo {author} {\bibfnamefont {J.}~\bibnamefont {Zhu}}, \bibinfo {author} {\bibfnamefont {P.}~\bibnamefont {Kn{\"u}ppel}}, \bibinfo {author} {\bibfnamefont {C.}~\bibnamefont {Vaswani}}, \bibinfo {author} {\bibfnamefont {K.}~\bibnamefont {Watanabe}}, \bibinfo {author} {\bibfnamefont {T.}~\bibnamefont {Taniguchi}}, \bibinfo {author} {\bibfnamefont {K.~F.}\ \bibnamefont {Mak}},\ and\ \bibinfo {author} {\bibfnamefont {J.}~\bibnamefont {Shan}},\ }\bibfield  {title} {\bibinfo {title} {Thermodynamic evidence of fractional chern insulator in moir{\'e} mote2},\ }\href {https://www.nature.com/articles/s41586-023-06452-3} {\bibfield  {journal} {\bibinfo  {journal} {Nature}\ }\textbf {\bibinfo {volume} {622}},\ \bibinfo {pages} {69} (\bibinfo {year} {2023})}\BibitemShut {NoStop}%
\bibitem [{\citenamefont {Kang}\ \emph {et~al.}(2024)\citenamefont {Kang}, \citenamefont {Shen}, \citenamefont {Qiu}, \citenamefont {Zeng}, \citenamefont {Xia}, \citenamefont {Watanabe}, \citenamefont {Taniguchi}, \citenamefont {Shan},\ and\ \citenamefont {Mak}}]{kang2024evidence}%
  \BibitemOpen
  \bibfield  {author} {\bibinfo {author} {\bibfnamefont {K.}~\bibnamefont {Kang}}, \bibinfo {author} {\bibfnamefont {B.}~\bibnamefont {Shen}}, \bibinfo {author} {\bibfnamefont {Y.}~\bibnamefont {Qiu}}, \bibinfo {author} {\bibfnamefont {Y.}~\bibnamefont {Zeng}}, \bibinfo {author} {\bibfnamefont {Z.}~\bibnamefont {Xia}}, \bibinfo {author} {\bibfnamefont {K.}~\bibnamefont {Watanabe}}, \bibinfo {author} {\bibfnamefont {T.}~\bibnamefont {Taniguchi}}, \bibinfo {author} {\bibfnamefont {J.}~\bibnamefont {Shan}},\ and\ \bibinfo {author} {\bibfnamefont {K.~F.}\ \bibnamefont {Mak}},\ }\bibfield  {title} {\bibinfo {title} {Evidence of the fractional quantum spin hall effect in moir{\'e} mote2},\ }\href {https://www.nature.com/articles/s41586-024-07214-5} {\bibfield  {journal} {\bibinfo  {journal} {Nature}\ }\textbf {\bibinfo {volume} {628}},\ \bibinfo {pages} {522} (\bibinfo {year} {2024})}\BibitemShut {NoStop}%
\bibitem [{\citenamefont {Lu}\ \emph {et~al.}(2024)\citenamefont {Lu}, \citenamefont {Han}, \citenamefont {Yao}, \citenamefont {Reddy}, \citenamefont {Yang}, \citenamefont {Seo}, \citenamefont {Watanabe}, \citenamefont {Taniguchi}, \citenamefont {Fu},\ and\ \citenamefont {Ju}}]{lu2024fractional}%
  \BibitemOpen
  \bibfield  {author} {\bibinfo {author} {\bibfnamefont {Z.}~\bibnamefont {Lu}}, \bibinfo {author} {\bibfnamefont {T.}~\bibnamefont {Han}}, \bibinfo {author} {\bibfnamefont {Y.}~\bibnamefont {Yao}}, \bibinfo {author} {\bibfnamefont {A.~P.}\ \bibnamefont {Reddy}}, \bibinfo {author} {\bibfnamefont {J.}~\bibnamefont {Yang}}, \bibinfo {author} {\bibfnamefont {J.}~\bibnamefont {Seo}}, \bibinfo {author} {\bibfnamefont {K.}~\bibnamefont {Watanabe}}, \bibinfo {author} {\bibfnamefont {T.}~\bibnamefont {Taniguchi}}, \bibinfo {author} {\bibfnamefont {L.}~\bibnamefont {Fu}},\ and\ \bibinfo {author} {\bibfnamefont {L.}~\bibnamefont {Ju}},\ }\bibfield  {title} {\bibinfo {title} {Fractional quantum anomalous hall effect in multilayer graphene},\ }\href {https://www.nature.com/articles/s41586-023-07010-7} {\bibfield  {journal} {\bibinfo  {journal} {Nature}\ }\textbf {\bibinfo {volume} {626}},\ \bibinfo {pages} {759} (\bibinfo {year} {2024})}\BibitemShut {NoStop}%
\bibitem [{\citenamefont {Xie}\ \emph {et~al.}(2024)\citenamefont {Xie}, \citenamefont {Huo}, \citenamefont {Lu}, \citenamefont {Feng}, \citenamefont {Zhang}, \citenamefont {Wang}, \citenamefont {Yang}, \citenamefont {Watanabe}, \citenamefont {Taniguchi}, \citenamefont {Liu}, \citenamefont {Song}, \citenamefont {Xie}, \citenamefont {Liu},\ and\ \citenamefont {Lu}}]{Xie_Huo_Lu_2024}%
  \BibitemOpen
  \bibfield  {author} {\bibinfo {author} {\bibfnamefont {J.}~\bibnamefont {Xie}}, \bibinfo {author} {\bibfnamefont {Z.}~\bibnamefont {Huo}}, \bibinfo {author} {\bibfnamefont {X.}~\bibnamefont {Lu}}, \bibinfo {author} {\bibfnamefont {Z.}~\bibnamefont {Feng}}, \bibinfo {author} {\bibfnamefont {Z.}~\bibnamefont {Zhang}}, \bibinfo {author} {\bibfnamefont {W.}~\bibnamefont {Wang}}, \bibinfo {author} {\bibfnamefont {Q.}~\bibnamefont {Yang}}, \bibinfo {author} {\bibfnamefont {K.}~\bibnamefont {Watanabe}}, \bibinfo {author} {\bibfnamefont {T.}~\bibnamefont {Taniguchi}}, \bibinfo {author} {\bibfnamefont {K.}~\bibnamefont {Liu}}, \bibinfo {author} {\bibfnamefont {Z.}~\bibnamefont {Song}}, \bibinfo {author} {\bibfnamefont {X.~C.}\ \bibnamefont {Xie}}, \bibinfo {author} {\bibfnamefont {J.}~\bibnamefont {Liu}},\ and\ \bibinfo {author} {\bibfnamefont {X.}~\bibnamefont {Lu}},\ }\bibfield  {title} {\bibinfo {title} {Even- and odd-denominator fractional quantum anomalous hall effect in graphene moire superlattices},\
  }\bibfield  {journal} {\bibinfo  {journal} {arXiv}\ }\href {https://doi.org/10.48550/arXiv.2405.16944} {10.48550/arXiv.2405.16944} (\bibinfo {year} {2024}),\ \bibinfo {note} {arXiv:2405.16944 [cond-mat]}\BibitemShut {NoStop}%
\bibitem [{\citenamefont {Li}\ \emph {et~al.}(2021)\citenamefont {Li}, \citenamefont {Kumar}, \citenamefont {Sun},\ and\ \citenamefont {Lin}}]{PhysRevResearch.3.L032070}%
  \BibitemOpen
  \bibfield  {author} {\bibinfo {author} {\bibfnamefont {H.}~\bibnamefont {Li}}, \bibinfo {author} {\bibfnamefont {U.}~\bibnamefont {Kumar}}, \bibinfo {author} {\bibfnamefont {K.}~\bibnamefont {Sun}},\ and\ \bibinfo {author} {\bibfnamefont {S.-Z.}\ \bibnamefont {Lin}},\ }\bibfield  {title} {\bibinfo {title} {Spontaneous fractional chern insulators in transition metal dichalcogenide moir\'e superlattices},\ }\href {https://doi.org/10.1103/PhysRevResearch.3.L032070} {\bibfield  {journal} {\bibinfo  {journal} {Phys. Rev. Res.}\ }\textbf {\bibinfo {volume} {3}},\ \bibinfo {pages} {L032070} (\bibinfo {year} {2021})}\BibitemShut {NoStop}%
\bibitem [{\citenamefont {Cr\'epel}\ and\ \citenamefont {Fu}(2023)}]{PhysRevB.107.L201109}%
  \BibitemOpen
  \bibfield  {author} {\bibinfo {author} {\bibfnamefont {V.}~\bibnamefont {Cr\'epel}}\ and\ \bibinfo {author} {\bibfnamefont {L.}~\bibnamefont {Fu}},\ }\bibfield  {title} {\bibinfo {title} {Anomalous hall metal and fractional chern insulator in twisted transition metal dichalcogenides},\ }\href {https://doi.org/10.1103/PhysRevB.107.L201109} {\bibfield  {journal} {\bibinfo  {journal} {Phys. Rev. B}\ }\textbf {\bibinfo {volume} {107}},\ \bibinfo {pages} {L201109} (\bibinfo {year} {2023})}\BibitemShut {NoStop}%
\bibitem [{\citenamefont {Provost}\ and\ \citenamefont {Vallee}(1980)}]{provost1980riemannian}%
  \BibitemOpen
  \bibfield  {author} {\bibinfo {author} {\bibfnamefont {J.}~\bibnamefont {Provost}}\ and\ \bibinfo {author} {\bibfnamefont {G.}~\bibnamefont {Vallee}},\ }\bibfield  {title} {\bibinfo {title} {Riemannian structure on manifolds of quantum states},\ }\href {https://link.springer.com/article/10.1007/bf02193559} {\bibfield  {journal} {\bibinfo  {journal} {Comm. Math. Phys.}\ }\textbf {\bibinfo {volume} {76}},\ \bibinfo {pages} {289} (\bibinfo {year} {1980})}\BibitemShut {NoStop}%
\bibitem [{\citenamefont {Klitzing}\ \emph {et~al.}(1980)\citenamefont {Klitzing}, \citenamefont {Dorda},\ and\ \citenamefont {Pepper}}]{PhysRevLett.45.494}%
  \BibitemOpen
  \bibfield  {author} {\bibinfo {author} {\bibfnamefont {K.~v.}\ \bibnamefont {Klitzing}}, \bibinfo {author} {\bibfnamefont {G.}~\bibnamefont {Dorda}},\ and\ \bibinfo {author} {\bibfnamefont {M.}~\bibnamefont {Pepper}},\ }\bibfield  {title} {\bibinfo {title} {New method for high-accuracy determination of the fine-structure constant based on quantized hall resistance},\ }\href {https://doi.org/10.1103/PhysRevLett.45.494} {\bibfield  {journal} {\bibinfo  {journal} {Phys. Rev. Lett.}\ }\textbf {\bibinfo {volume} {45}},\ \bibinfo {pages} {494} (\bibinfo {year} {1980})}\BibitemShut {NoStop}%
\bibitem [{\citenamefont {Thouless}\ \emph {et~al.}(1982)\citenamefont {Thouless}, \citenamefont {Kohmoto}, \citenamefont {Nightingale},\ and\ \citenamefont {den Nijs}}]{PhysRevLett.49.405}%
  \BibitemOpen
  \bibfield  {author} {\bibinfo {author} {\bibfnamefont {D.~J.}\ \bibnamefont {Thouless}}, \bibinfo {author} {\bibfnamefont {M.}~\bibnamefont {Kohmoto}}, \bibinfo {author} {\bibfnamefont {M.~P.}\ \bibnamefont {Nightingale}},\ and\ \bibinfo {author} {\bibfnamefont {M.}~\bibnamefont {den Nijs}},\ }\bibfield  {title} {\bibinfo {title} {Quantized hall conductance in a two-dimensional periodic potential},\ }\href {https://doi.org/10.1103/PhysRevLett.49.405} {\bibfield  {journal} {\bibinfo  {journal} {Phys. Rev. Lett.}\ }\textbf {\bibinfo {volume} {49}},\ \bibinfo {pages} {405} (\bibinfo {year} {1982})}\BibitemShut {NoStop}%
\bibitem [{\citenamefont {Haldane}(1988)}]{PhysRevLett.61.2015}%
  \BibitemOpen
  \bibfield  {author} {\bibinfo {author} {\bibfnamefont {F.~D.~M.}\ \bibnamefont {Haldane}},\ }\bibfield  {title} {\bibinfo {title} {Model for a quantum hall effect without landau levels: Condensed-matter realization of the "parity anomaly"},\ }\href {https://doi.org/10.1103/PhysRevLett.61.2015} {\bibfield  {journal} {\bibinfo  {journal} {Phys. Rev. Lett.}\ }\textbf {\bibinfo {volume} {61}},\ \bibinfo {pages} {2015} (\bibinfo {year} {1988})}\BibitemShut {NoStop}%
\bibitem [{\citenamefont {Kane}\ and\ \citenamefont {Mele}(2005)}]{PhysRevLett.95.146802}%
  \BibitemOpen
  \bibfield  {author} {\bibinfo {author} {\bibfnamefont {C.~L.}\ \bibnamefont {Kane}}\ and\ \bibinfo {author} {\bibfnamefont {E.~J.}\ \bibnamefont {Mele}},\ }\bibfield  {title} {\bibinfo {title} {${Z}_{2}$ topological order and the quantum spin hall effect},\ }\href {https://doi.org/10.1103/PhysRevLett.95.146802} {\bibfield  {journal} {\bibinfo  {journal} {Phys. Rev. Lett.}\ }\textbf {\bibinfo {volume} {95}},\ \bibinfo {pages} {146802} (\bibinfo {year} {2005})}\BibitemShut {NoStop}%
\bibitem [{\citenamefont {Wan}\ \emph {et~al.}(2011)\citenamefont {Wan}, \citenamefont {Turner}, \citenamefont {Vishwanath},\ and\ \citenamefont {Savrasov}}]{PhysRevB.83.205101}%
  \BibitemOpen
  \bibfield  {author} {\bibinfo {author} {\bibfnamefont {X.}~\bibnamefont {Wan}}, \bibinfo {author} {\bibfnamefont {A.~M.}\ \bibnamefont {Turner}}, \bibinfo {author} {\bibfnamefont {A.}~\bibnamefont {Vishwanath}},\ and\ \bibinfo {author} {\bibfnamefont {S.~Y.}\ \bibnamefont {Savrasov}},\ }\bibfield  {title} {\bibinfo {title} {Topological semimetal and fermi-arc surface states in the electronic structure of pyrochlore iridates},\ }\href {https://doi.org/10.1103/PhysRevB.83.205101} {\bibfield  {journal} {\bibinfo  {journal} {Phys. Rev. B}\ }\textbf {\bibinfo {volume} {83}},\ \bibinfo {pages} {205101} (\bibinfo {year} {2011})}\BibitemShut {NoStop}%
\bibitem [{\citenamefont {Wang}\ \emph {et~al.}(2021{\natexlab{a}})\citenamefont {Wang}, \citenamefont {Gao},\ and\ \citenamefont {Xiao}}]{PhysRevLett.127.277201}%
  \BibitemOpen
  \bibfield  {author} {\bibinfo {author} {\bibfnamefont {C.}~\bibnamefont {Wang}}, \bibinfo {author} {\bibfnamefont {Y.}~\bibnamefont {Gao}},\ and\ \bibinfo {author} {\bibfnamefont {D.}~\bibnamefont {Xiao}},\ }\bibfield  {title} {\bibinfo {title} {Intrinsic nonlinear hall effect in antiferromagnetic tetragonal cumnas},\ }\href {https://doi.org/10.1103/PhysRevLett.127.277201} {\bibfield  {journal} {\bibinfo  {journal} {Phys. Rev. Lett.}\ }\textbf {\bibinfo {volume} {127}},\ \bibinfo {pages} {277201} (\bibinfo {year} {2021}{\natexlab{a}})}\BibitemShut {NoStop}%
\bibitem [{\citenamefont {Gao}\ \emph {et~al.}(2014)\citenamefont {Gao}, \citenamefont {Yang},\ and\ \citenamefont {Niu}}]{PhysRevLett.112.166601}%
  \BibitemOpen
  \bibfield  {author} {\bibinfo {author} {\bibfnamefont {Y.}~\bibnamefont {Gao}}, \bibinfo {author} {\bibfnamefont {S.~A.}\ \bibnamefont {Yang}},\ and\ \bibinfo {author} {\bibfnamefont {Q.}~\bibnamefont {Niu}},\ }\bibfield  {title} {\bibinfo {title} {Field induced positional shift of bloch electrons and its dynamical implications},\ }\href {https://doi.org/10.1103/PhysRevLett.112.166601} {\bibfield  {journal} {\bibinfo  {journal} {Phys. Rev. Lett.}\ }\textbf {\bibinfo {volume} {112}},\ \bibinfo {pages} {166601} (\bibinfo {year} {2014})}\BibitemShut {NoStop}%
\bibitem [{\citenamefont {Kaplan}\ \emph {et~al.}(2024)\citenamefont {Kaplan}, \citenamefont {Holder},\ and\ \citenamefont {Yan}}]{PhysRevLett.132.026301}%
  \BibitemOpen
  \bibfield  {author} {\bibinfo {author} {\bibfnamefont {D.}~\bibnamefont {Kaplan}}, \bibinfo {author} {\bibfnamefont {T.}~\bibnamefont {Holder}},\ and\ \bibinfo {author} {\bibfnamefont {B.}~\bibnamefont {Yan}},\ }\bibfield  {title} {\bibinfo {title} {Unification of nonlinear anomalous hall effect and nonreciprocal magnetoresistance in metals by the quantum geometry},\ }\href {https://doi.org/10.1103/PhysRevLett.132.026301} {\bibfield  {journal} {\bibinfo  {journal} {Phys. Rev. Lett.}\ }\textbf {\bibinfo {volume} {132}},\ \bibinfo {pages} {026301} (\bibinfo {year} {2024})}\BibitemShut {NoStop}%
\bibitem [{\citenamefont {Gao}\ \emph {et~al.}(2023)\citenamefont {Gao}, \citenamefont {Liu}, \citenamefont {Qiu}, \citenamefont {Ghosh}, \citenamefont {Trevisan}, \citenamefont {Onishi}, \citenamefont {Hu}, \citenamefont {Qian}, \citenamefont {Tien}, \citenamefont {Chen}, \citenamefont {Huang}, \citenamefont {Bérubé}, \citenamefont {Li}, \citenamefont {Tzschaschel}, \citenamefont {Dinh}, \citenamefont {Sun}, \citenamefont {Ho}, \citenamefont {Lien}, \citenamefont {Singh}, \citenamefont {Watanabe}, \citenamefont {Taniguchi}, \citenamefont {Bell}, \citenamefont {Lin}, \citenamefont {Chang}, \citenamefont {Du}, \citenamefont {Bansil}, \citenamefont {Fu}, \citenamefont {Ni}, \citenamefont {Orth}, \citenamefont {Ma},\ and\ \citenamefont {Xu}}]{Gao_Liu_Qiu_Ghosh2023}%
  \BibitemOpen
  \bibfield  {author} {\bibinfo {author} {\bibfnamefont {A.}~\bibnamefont {Gao}}, \bibinfo {author} {\bibfnamefont {Y.-F.}\ \bibnamefont {Liu}}, \bibinfo {author} {\bibfnamefont {J.-X.}\ \bibnamefont {Qiu}}, \bibinfo {author} {\bibfnamefont {B.}~\bibnamefont {Ghosh}}, \bibinfo {author} {\bibfnamefont {T.~V.}\ \bibnamefont {Trevisan}}, \bibinfo {author} {\bibfnamefont {Y.}~\bibnamefont {Onishi}}, \bibinfo {author} {\bibfnamefont {C.}~\bibnamefont {Hu}}, \bibinfo {author} {\bibfnamefont {T.}~\bibnamefont {Qian}}, \bibinfo {author} {\bibfnamefont {H.-J.}\ \bibnamefont {Tien}}, \bibinfo {author} {\bibfnamefont {S.-W.}\ \bibnamefont {Chen}}, \bibinfo {author} {\bibfnamefont {M.}~\bibnamefont {Huang}}, \bibinfo {author} {\bibfnamefont {D.}~\bibnamefont {Bérubé}}, \bibinfo {author} {\bibfnamefont {H.}~\bibnamefont {Li}}, \bibinfo {author} {\bibfnamefont {C.}~\bibnamefont {Tzschaschel}}, \bibinfo {author} {\bibfnamefont {T.}~\bibnamefont {Dinh}}, \bibinfo {author} {\bibfnamefont {Z.}~\bibnamefont {Sun}}, \bibinfo
  {author} {\bibfnamefont {S.-C.}\ \bibnamefont {Ho}}, \bibinfo {author} {\bibfnamefont {S.-W.}\ \bibnamefont {Lien}}, \bibinfo {author} {\bibfnamefont {B.}~\bibnamefont {Singh}}, \bibinfo {author} {\bibfnamefont {K.}~\bibnamefont {Watanabe}}, \bibinfo {author} {\bibfnamefont {T.}~\bibnamefont {Taniguchi}}, \bibinfo {author} {\bibfnamefont {D.~C.}\ \bibnamefont {Bell}}, \bibinfo {author} {\bibfnamefont {H.}~\bibnamefont {Lin}}, \bibinfo {author} {\bibfnamefont {T.-R.}\ \bibnamefont {Chang}}, \bibinfo {author} {\bibfnamefont {C.~R.}\ \bibnamefont {Du}}, \bibinfo {author} {\bibfnamefont {A.}~\bibnamefont {Bansil}}, \bibinfo {author} {\bibfnamefont {L.}~\bibnamefont {Fu}}, \bibinfo {author} {\bibfnamefont {N.}~\bibnamefont {Ni}}, \bibinfo {author} {\bibfnamefont {P.~P.}\ \bibnamefont {Orth}}, \bibinfo {author} {\bibfnamefont {Q.}~\bibnamefont {Ma}},\ and\ \bibinfo {author} {\bibfnamefont {S.-Y.}\ \bibnamefont {Xu}},\ }\bibfield  {title} {\bibinfo {title} {Quantum metric nonlinear hall effect in a topological
  antiferromagnetic heterostructure},\ }\href {https://doi.org/10.1126/science.adf1506} {\bibfield  {journal} {\bibinfo  {journal} {Science}\ }\textbf {\bibinfo {volume} {381}},\ \bibinfo {pages} {181} (\bibinfo {year} {2023})}\BibitemShut {NoStop}%
\bibitem [{\citenamefont {Gao}\ and\ \citenamefont {Xiao}(2019)}]{PhysRevLett.122.227402}%
  \BibitemOpen
  \bibfield  {author} {\bibinfo {author} {\bibfnamefont {Y.}~\bibnamefont {Gao}}\ and\ \bibinfo {author} {\bibfnamefont {D.}~\bibnamefont {Xiao}},\ }\bibfield  {title} {\bibinfo {title} {Nonreciprocal directional dichroism induced by the quantum metric dipole},\ }\href {https://doi.org/10.1103/PhysRevLett.122.227402} {\bibfield  {journal} {\bibinfo  {journal} {Phys. Rev. Lett.}\ }\textbf {\bibinfo {volume} {122}},\ \bibinfo {pages} {227402} (\bibinfo {year} {2019})}\BibitemShut {NoStop}%
\bibitem [{\citenamefont {Peotta}\ and\ \citenamefont {T{\"o}rm{\"a}}(2015)}]{peotta2015superfluidity}%
  \BibitemOpen
  \bibfield  {author} {\bibinfo {author} {\bibfnamefont {S.}~\bibnamefont {Peotta}}\ and\ \bibinfo {author} {\bibfnamefont {P.}~\bibnamefont {T{\"o}rm{\"a}}},\ }\bibfield  {title} {\bibinfo {title} {Superfluidity in topologically nontrivial flat bands},\ }\href {https://www.nature.com/articles/ncomms9944} {\bibfield  {journal} {\bibinfo  {journal} {Nature communications}\ }\textbf {\bibinfo {volume} {6}},\ \bibinfo {pages} {8944} (\bibinfo {year} {2015})}\BibitemShut {NoStop}%
\bibitem [{\citenamefont {T{\"o}rm{\"a}}\ \emph {et~al.}(2022)\citenamefont {T{\"o}rm{\"a}}, \citenamefont {Peotta},\ and\ \citenamefont {Bernevig}}]{torma2022superconductivity}%
  \BibitemOpen
  \bibfield  {author} {\bibinfo {author} {\bibfnamefont {P.}~\bibnamefont {T{\"o}rm{\"a}}}, \bibinfo {author} {\bibfnamefont {S.}~\bibnamefont {Peotta}},\ and\ \bibinfo {author} {\bibfnamefont {B.~A.}\ \bibnamefont {Bernevig}},\ }\bibfield  {title} {\bibinfo {title} {Superconductivity, superfluidity and quantum geometry in twisted multilayer systems},\ }\href {https://www.nature.com/articles/s42254-022-00466-y} {\bibfield  {journal} {\bibinfo  {journal} {Nature Reviews Physics}\ }\textbf {\bibinfo {volume} {4}},\ \bibinfo {pages} {528} (\bibinfo {year} {2022})}\BibitemShut {NoStop}%
\bibitem [{\citenamefont {Parameswaran}\ \emph {et~al.}(2013)\citenamefont {Parameswaran}, \citenamefont {Roy},\ and\ \citenamefont {Sondhi}}]{parameswaran2013fractional}%
  \BibitemOpen
  \bibfield  {author} {\bibinfo {author} {\bibfnamefont {S.~A.}\ \bibnamefont {Parameswaran}}, \bibinfo {author} {\bibfnamefont {R.}~\bibnamefont {Roy}},\ and\ \bibinfo {author} {\bibfnamefont {S.~L.}\ \bibnamefont {Sondhi}},\ }\bibfield  {title} {\bibinfo {title} {Fractional quantum hall physics in topological flat bands},\ }\href {https://www.sciencedirect.com/science/article/pii/S163107051300073X} {\bibfield  {journal} {\bibinfo  {journal} {Comptes Rendus Physique}\ }\textbf {\bibinfo {volume} {14}},\ \bibinfo {pages} {816} (\bibinfo {year} {2013})}\BibitemShut {NoStop}%
\bibitem [{\citenamefont {BERGHOLTZ}\ and\ \citenamefont {LIU}(2013)}]{Bergholtz_Liu_2013}%
  \BibitemOpen
  \bibfield  {author} {\bibinfo {author} {\bibfnamefont {E.~J.}\ \bibnamefont {BERGHOLTZ}}\ and\ \bibinfo {author} {\bibfnamefont {Z.}~\bibnamefont {LIU}},\ }\bibfield  {title} {\bibinfo {title} {Topological flat band models and fractional chern insulators},\ }\href {https://doi.org/10.1142/S021797921330017X} {\bibfield  {journal} {\bibinfo  {journal} {International Journal of Modern Physics B}\ }\textbf {\bibinfo {volume} {27}},\ \bibinfo {pages} {1330017} (\bibinfo {year} {2013})}\BibitemShut {NoStop}%
\bibitem [{\citenamefont {Neupert}\ \emph {et~al.}(2015)\citenamefont {Neupert}, \citenamefont {Chamon}, \citenamefont {Iadecola}, \citenamefont {Santos},\ and\ \citenamefont {Mudry}}]{neupert2015fractional}%
  \BibitemOpen
  \bibfield  {author} {\bibinfo {author} {\bibfnamefont {T.}~\bibnamefont {Neupert}}, \bibinfo {author} {\bibfnamefont {C.}~\bibnamefont {Chamon}}, \bibinfo {author} {\bibfnamefont {T.}~\bibnamefont {Iadecola}}, \bibinfo {author} {\bibfnamefont {L.~H.}\ \bibnamefont {Santos}},\ and\ \bibinfo {author} {\bibfnamefont {C.}~\bibnamefont {Mudry}},\ }\bibfield  {title} {\bibinfo {title} {Fractional (chern and topological) insulators},\ }\href {https://dx.doi.org/10.1088/0031-8949/2015/T164/014005} {\bibfield  {journal} {\bibinfo  {journal} {Physica Scripta}\ }\textbf {\bibinfo {volume} {2015}},\ \bibinfo {pages} {014005} (\bibinfo {year} {2015})}\BibitemShut {NoStop}%
\bibitem [{\citenamefont {Liu}\ and\ \citenamefont {Bergholtz}(2024)}]{LIU2024515}%
  \BibitemOpen
  \bibfield  {author} {\bibinfo {author} {\bibfnamefont {Z.}~\bibnamefont {Liu}}\ and\ \bibinfo {author} {\bibfnamefont {E.~J.}\ \bibnamefont {Bergholtz}},\ }\bibfield  {title} {\bibinfo {title} {Recent developments in fractional chern insulators},\ }\href {https://doi.org/https://doi.org/10.1016/B978-0-323-90800-9.00136-0} {\bibfield  {journal} {\bibinfo  {journal} {Encyclopedia of Condensed Matter Physics}\ ,\ \bibinfo {pages} {515}} (\bibinfo {year} {2024})}\BibitemShut {NoStop}%
\bibitem [{\citenamefont {Morales-Dur\'an}\ \emph {et~al.}(2023)\citenamefont {Morales-Dur\'an}, \citenamefont {Wang}, \citenamefont {Schleder}, \citenamefont {Angeli}, \citenamefont {Zhu}, \citenamefont {Kaxiras}, \citenamefont {Repellin},\ and\ \citenamefont {Cano}}]{PhysRevResearch.5.L032022}%
  \BibitemOpen
  \bibfield  {author} {\bibinfo {author} {\bibfnamefont {N.}~\bibnamefont {Morales-Dur\'an}}, \bibinfo {author} {\bibfnamefont {J.}~\bibnamefont {Wang}}, \bibinfo {author} {\bibfnamefont {G.~R.}\ \bibnamefont {Schleder}}, \bibinfo {author} {\bibfnamefont {M.}~\bibnamefont {Angeli}}, \bibinfo {author} {\bibfnamefont {Z.}~\bibnamefont {Zhu}}, \bibinfo {author} {\bibfnamefont {E.}~\bibnamefont {Kaxiras}}, \bibinfo {author} {\bibfnamefont {C.}~\bibnamefont {Repellin}},\ and\ \bibinfo {author} {\bibfnamefont {J.}~\bibnamefont {Cano}},\ }\bibfield  {title} {\bibinfo {title} {Pressure-enhanced fractional chern insulators along a magic line in moir\'e transition metal dichalcogenides},\ }\href {https://doi.org/10.1103/PhysRevResearch.5.L032022} {\bibfield  {journal} {\bibinfo  {journal} {Phys. Rev. Res.}\ }\textbf {\bibinfo {volume} {5}},\ \bibinfo {pages} {L032022} (\bibinfo {year} {2023})}\BibitemShut {NoStop}%
\bibitem [{\citenamefont {Morales-Dur\'an}\ \emph {et~al.}(2024)\citenamefont {Morales-Dur\'an}, \citenamefont {Wei}, \citenamefont {Shi},\ and\ \citenamefont {MacDonald}}]{PhysRevLett.132.096602}%
  \BibitemOpen
  \bibfield  {author} {\bibinfo {author} {\bibfnamefont {N.}~\bibnamefont {Morales-Dur\'an}}, \bibinfo {author} {\bibfnamefont {N.}~\bibnamefont {Wei}}, \bibinfo {author} {\bibfnamefont {J.}~\bibnamefont {Shi}},\ and\ \bibinfo {author} {\bibfnamefont {A.~H.}\ \bibnamefont {MacDonald}},\ }\bibfield  {title} {\bibinfo {title} {Magic angles and fractional chern insulators in twisted homobilayer transition metal dichalcogenides},\ }\href {https://doi.org/10.1103/PhysRevLett.132.096602} {\bibfield  {journal} {\bibinfo  {journal} {Phys. Rev. Lett.}\ }\textbf {\bibinfo {volume} {132}},\ \bibinfo {pages} {096602} (\bibinfo {year} {2024})}\BibitemShut {NoStop}%
\bibitem [{\citenamefont {Shi}\ \emph {et~al.}(2024)\citenamefont {Shi}, \citenamefont {Morales-Dur\'an}, \citenamefont {Khalaf},\ and\ \citenamefont {MacDonald}}]{PhysRevB.110.035130}%
  \BibitemOpen
  \bibfield  {author} {\bibinfo {author} {\bibfnamefont {J.}~\bibnamefont {Shi}}, \bibinfo {author} {\bibfnamefont {N.}~\bibnamefont {Morales-Dur\'an}}, \bibinfo {author} {\bibfnamefont {E.}~\bibnamefont {Khalaf}},\ and\ \bibinfo {author} {\bibfnamefont {A.~H.}\ \bibnamefont {MacDonald}},\ }\bibfield  {title} {\bibinfo {title} {Adiabatic approximation and aharonov-casher bands in twisted homobilayer transition metal dichalcogenides},\ }\href {https://doi.org/10.1103/PhysRevB.110.035130} {\bibfield  {journal} {\bibinfo  {journal} {Phys. Rev. B}\ }\textbf {\bibinfo {volume} {110}},\ \bibinfo {pages} {035130} (\bibinfo {year} {2024})}\BibitemShut {NoStop}%
\bibitem [{\citenamefont {Shavit}\ and\ \citenamefont {Oreg}(2024)}]{shavit2024quantum}%
  \BibitemOpen
  \bibfield  {author} {\bibinfo {author} {\bibfnamefont {G.}~\bibnamefont {Shavit}}\ and\ \bibinfo {author} {\bibfnamefont {Y.}~\bibnamefont {Oreg}},\ }\bibfield  {title} {\bibinfo {title} {Quantum geometry and stabilization of fractional chern insulators far from the ideal limit},\ }\href {https://arxiv.org/abs/2405.09627} {\bibfield  {journal} {\bibinfo  {journal} {arXiv:2405.09627}\ } (\bibinfo {year} {2024})}\BibitemShut {NoStop}%
\bibitem [{\citenamefont {Tang}\ \emph {et~al.}(2011)\citenamefont {Tang}, \citenamefont {Mei},\ and\ \citenamefont {Wen}}]{PhysRevLett.106.236802}%
  \BibitemOpen
  \bibfield  {author} {\bibinfo {author} {\bibfnamefont {E.}~\bibnamefont {Tang}}, \bibinfo {author} {\bibfnamefont {J.-W.}\ \bibnamefont {Mei}},\ and\ \bibinfo {author} {\bibfnamefont {X.-G.}\ \bibnamefont {Wen}},\ }\bibfield  {title} {\bibinfo {title} {High-temperature fractional quantum hall states},\ }\href {https://doi.org/10.1103/PhysRevLett.106.236802} {\bibfield  {journal} {\bibinfo  {journal} {Phys. Rev. Lett.}\ }\textbf {\bibinfo {volume} {106}},\ \bibinfo {pages} {236802} (\bibinfo {year} {2011})}\BibitemShut {NoStop}%
\bibitem [{\citenamefont {Sun}\ \emph {et~al.}(2011)\citenamefont {Sun}, \citenamefont {Gu}, \citenamefont {Katsura},\ and\ \citenamefont {Das~Sarma}}]{PhysRevLett.106.236803}%
  \BibitemOpen
  \bibfield  {author} {\bibinfo {author} {\bibfnamefont {K.}~\bibnamefont {Sun}}, \bibinfo {author} {\bibfnamefont {Z.}~\bibnamefont {Gu}}, \bibinfo {author} {\bibfnamefont {H.}~\bibnamefont {Katsura}},\ and\ \bibinfo {author} {\bibfnamefont {S.}~\bibnamefont {Das~Sarma}},\ }\bibfield  {title} {\bibinfo {title} {Nearly flatbands with nontrivial topology},\ }\href {https://doi.org/10.1103/PhysRevLett.106.236803} {\bibfield  {journal} {\bibinfo  {journal} {Phys. Rev. Lett.}\ }\textbf {\bibinfo {volume} {106}},\ \bibinfo {pages} {236803} (\bibinfo {year} {2011})}\BibitemShut {NoStop}%
\bibitem [{\citenamefont {Neupert}\ \emph {et~al.}(2011)\citenamefont {Neupert}, \citenamefont {Santos}, \citenamefont {Chamon},\ and\ \citenamefont {Mudry}}]{PhysRevLett.106.236804}%
  \BibitemOpen
  \bibfield  {author} {\bibinfo {author} {\bibfnamefont {T.}~\bibnamefont {Neupert}}, \bibinfo {author} {\bibfnamefont {L.}~\bibnamefont {Santos}}, \bibinfo {author} {\bibfnamefont {C.}~\bibnamefont {Chamon}},\ and\ \bibinfo {author} {\bibfnamefont {C.}~\bibnamefont {Mudry}},\ }\bibfield  {title} {\bibinfo {title} {Fractional quantum hall states at zero magnetic field},\ }\href {https://doi.org/10.1103/PhysRevLett.106.236804} {\bibfield  {journal} {\bibinfo  {journal} {Phys. Rev. Lett.}\ }\textbf {\bibinfo {volume} {106}},\ \bibinfo {pages} {236804} (\bibinfo {year} {2011})}\BibitemShut {NoStop}%
\bibitem [{\citenamefont {Sheng}\ \emph {et~al.}(2011)\citenamefont {Sheng}, \citenamefont {Gu}, \citenamefont {Sun},\ and\ \citenamefont {Sheng}}]{sheng2011fractional}%
  \BibitemOpen
  \bibfield  {author} {\bibinfo {author} {\bibfnamefont {D.~N.}\ \bibnamefont {Sheng}}, \bibinfo {author} {\bibfnamefont {Z.-C.}\ \bibnamefont {Gu}}, \bibinfo {author} {\bibfnamefont {K.}~\bibnamefont {Sun}},\ and\ \bibinfo {author} {\bibfnamefont {L.}~\bibnamefont {Sheng}},\ }\bibfield  {title} {\bibinfo {title} {Fractional quantum hall effect in the absence of landau levels},\ }\href {https://www.nature.com/articles/ncomms1380} {\bibfield  {journal} {\bibinfo  {journal} {Nat. Commun.}\ }\textbf {\bibinfo {volume} {2}},\ \bibinfo {pages} {389} (\bibinfo {year} {2011})}\BibitemShut {NoStop}%
\bibitem [{\citenamefont {Regnault}\ and\ \citenamefont {Bernevig}(2011)}]{PhysRevX.1.021014}%
  \BibitemOpen
  \bibfield  {author} {\bibinfo {author} {\bibfnamefont {N.}~\bibnamefont {Regnault}}\ and\ \bibinfo {author} {\bibfnamefont {B.~A.}\ \bibnamefont {Bernevig}},\ }\bibfield  {title} {\bibinfo {title} {Fractional chern insulator},\ }\href {https://doi.org/10.1103/PhysRevX.1.021014} {\bibfield  {journal} {\bibinfo  {journal} {Phys. Rev. X}\ }\textbf {\bibinfo {volume} {1}},\ \bibinfo {pages} {021014} (\bibinfo {year} {2011})}\BibitemShut {NoStop}%
\bibitem [{\citenamefont {Dong}\ \emph {et~al.}(2023)\citenamefont {Dong}, \citenamefont {Wang}, \citenamefont {Ledwith}, \citenamefont {Vishwanath},\ and\ \citenamefont {Parker}}]{PhysRevLett.131.136502}%
  \BibitemOpen
  \bibfield  {author} {\bibinfo {author} {\bibfnamefont {J.}~\bibnamefont {Dong}}, \bibinfo {author} {\bibfnamefont {J.}~\bibnamefont {Wang}}, \bibinfo {author} {\bibfnamefont {P.~J.}\ \bibnamefont {Ledwith}}, \bibinfo {author} {\bibfnamefont {A.}~\bibnamefont {Vishwanath}},\ and\ \bibinfo {author} {\bibfnamefont {D.~E.}\ \bibnamefont {Parker}},\ }\bibfield  {title} {\bibinfo {title} {Composite fermi liquid at zero magnetic field in twisted ${\mathrm{mote}}_{2}$},\ }\href {https://doi.org/10.1103/PhysRevLett.131.136502} {\bibfield  {journal} {\bibinfo  {journal} {Phys. Rev. Lett.}\ }\textbf {\bibinfo {volume} {131}},\ \bibinfo {pages} {136502} (\bibinfo {year} {2023})}\BibitemShut {NoStop}%
\bibitem [{\citenamefont {Het\'enyi}\ and\ \citenamefont {L\'evay}(2023)}]{PhysRevA.108.032218}%
  \BibitemOpen
  \bibfield  {author} {\bibinfo {author} {\bibfnamefont {B.}~\bibnamefont {Het\'enyi}}\ and\ \bibinfo {author} {\bibfnamefont {P.}~\bibnamefont {L\'evay}},\ }\bibfield  {title} {\bibinfo {title} {Fluctuations, uncertainty relations, and the geometry of quantum state manifolds},\ }\href {https://doi.org/10.1103/PhysRevA.108.032218} {\bibfield  {journal} {\bibinfo  {journal} {Phys. Rev. A}\ }\textbf {\bibinfo {volume} {108}},\ \bibinfo {pages} {032218} (\bibinfo {year} {2023})}\BibitemShut {NoStop}%
\bibitem [{\citenamefont {Roy}(2014)}]{PhysRevB.90.165139}%
  \BibitemOpen
  \bibfield  {author} {\bibinfo {author} {\bibfnamefont {R.}~\bibnamefont {Roy}},\ }\bibfield  {title} {\bibinfo {title} {Band geometry of fractional topological insulators},\ }\href {https://doi.org/10.1103/PhysRevB.90.165139} {\bibfield  {journal} {\bibinfo  {journal} {Phys. Rev. B}\ }\textbf {\bibinfo {volume} {90}},\ \bibinfo {pages} {165139} (\bibinfo {year} {2014})}\BibitemShut {NoStop}%
\bibitem [{\citenamefont {Ledwith}\ \emph {et~al.}(2020)\citenamefont {Ledwith}, \citenamefont {Tarnopolsky}, \citenamefont {Khalaf},\ and\ \citenamefont {Vishwanath}}]{PhysRevResearch.2.023237}%
  \BibitemOpen
  \bibfield  {author} {\bibinfo {author} {\bibfnamefont {P.~J.}\ \bibnamefont {Ledwith}}, \bibinfo {author} {\bibfnamefont {G.}~\bibnamefont {Tarnopolsky}}, \bibinfo {author} {\bibfnamefont {E.}~\bibnamefont {Khalaf}},\ and\ \bibinfo {author} {\bibfnamefont {A.}~\bibnamefont {Vishwanath}},\ }\bibfield  {title} {\bibinfo {title} {Fractional chern insulator states in twisted bilayer graphene: An analytical approach},\ }\href {https://doi.org/10.1103/PhysRevResearch.2.023237} {\bibfield  {journal} {\bibinfo  {journal} {Phys. Rev. Res.}\ }\textbf {\bibinfo {volume} {2}},\ \bibinfo {pages} {023237} (\bibinfo {year} {2020})}\BibitemShut {NoStop}%
\bibitem [{\citenamefont {Wang}\ \emph {et~al.}(2021{\natexlab{b}})\citenamefont {Wang}, \citenamefont {Cano}, \citenamefont {Millis}, \citenamefont {Liu},\ and\ \citenamefont {Yang}}]{PhysRevLett.127.246403}%
  \BibitemOpen
  \bibfield  {author} {\bibinfo {author} {\bibfnamefont {J.}~\bibnamefont {Wang}}, \bibinfo {author} {\bibfnamefont {J.}~\bibnamefont {Cano}}, \bibinfo {author} {\bibfnamefont {A.~J.}\ \bibnamefont {Millis}}, \bibinfo {author} {\bibfnamefont {Z.}~\bibnamefont {Liu}},\ and\ \bibinfo {author} {\bibfnamefont {B.}~\bibnamefont {Yang}},\ }\bibfield  {title} {\bibinfo {title} {Exact landau level description of geometry and interaction in a flatband},\ }\href {https://doi.org/10.1103/PhysRevLett.127.246403} {\bibfield  {journal} {\bibinfo  {journal} {Phys. Rev. Lett.}\ }\textbf {\bibinfo {volume} {127}},\ \bibinfo {pages} {246403} (\bibinfo {year} {2021}{\natexlab{b}})}\BibitemShut {NoStop}%
\bibitem [{\citenamefont {Wang}\ \emph {et~al.}(2023)\citenamefont {Wang}, \citenamefont {Klevtsov},\ and\ \citenamefont {Liu}}]{PhysRevResearch.5.023167}%
  \BibitemOpen
  \bibfield  {author} {\bibinfo {author} {\bibfnamefont {J.}~\bibnamefont {Wang}}, \bibinfo {author} {\bibfnamefont {S.}~\bibnamefont {Klevtsov}},\ and\ \bibinfo {author} {\bibfnamefont {Z.}~\bibnamefont {Liu}},\ }\bibfield  {title} {\bibinfo {title} {Origin of model fractional chern insulators in all topological ideal flatbands: Explicit color-entangled wave function and exact density algebra},\ }\href {https://doi.org/10.1103/PhysRevResearch.5.023167} {\bibfield  {journal} {\bibinfo  {journal} {Phys. Rev. Res.}\ }\textbf {\bibinfo {volume} {5}},\ \bibinfo {pages} {023167} (\bibinfo {year} {2023})}\BibitemShut {NoStop}%
\bibitem [{\citenamefont {Girvin}\ \emph {et~al.}(1986)\citenamefont {Girvin}, \citenamefont {MacDonald},\ and\ \citenamefont {Platzman}}]{PhysRevB.33.2481}%
  \BibitemOpen
  \bibfield  {author} {\bibinfo {author} {\bibfnamefont {S.~M.}\ \bibnamefont {Girvin}}, \bibinfo {author} {\bibfnamefont {A.~H.}\ \bibnamefont {MacDonald}},\ and\ \bibinfo {author} {\bibfnamefont {P.~M.}\ \bibnamefont {Platzman}},\ }\bibfield  {title} {\bibinfo {title} {Magneto-roton theory of collective excitations in the fractional quantum hall effect},\ }\href {https://doi.org/10.1103/PhysRevB.33.2481} {\bibfield  {journal} {\bibinfo  {journal} {Phys. Rev. B}\ }\textbf {\bibinfo {volume} {33}},\ \bibinfo {pages} {2481} (\bibinfo {year} {1986})}\BibitemShut {NoStop}%
\bibitem [{\citenamefont {Bernevig}\ \emph {et~al.}(2021)\citenamefont {Bernevig}, \citenamefont {Song}, \citenamefont {Regnault},\ and\ \citenamefont {Lian}}]{PhysRevB.103.205413}%
  \BibitemOpen
  \bibfield  {author} {\bibinfo {author} {\bibfnamefont {B.~A.}\ \bibnamefont {Bernevig}}, \bibinfo {author} {\bibfnamefont {Z.-D.}\ \bibnamefont {Song}}, \bibinfo {author} {\bibfnamefont {N.}~\bibnamefont {Regnault}},\ and\ \bibinfo {author} {\bibfnamefont {B.}~\bibnamefont {Lian}},\ }\bibfield  {title} {\bibinfo {title} {Twisted bilayer graphene. iii. interacting hamiltonian and exact symmetries},\ }\href {https://doi.org/10.1103/PhysRevB.103.205413} {\bibfield  {journal} {\bibinfo  {journal} {Phys. Rev. B}\ }\textbf {\bibinfo {volume} {103}},\ \bibinfo {pages} {205413} (\bibinfo {year} {2021})}\BibitemShut {NoStop}%
\bibitem [{\citenamefont {Ozawa}\ and\ \citenamefont {Mera}(2021)}]{PhysRevB.104.045103}%
  \BibitemOpen
  \bibfield  {author} {\bibinfo {author} {\bibfnamefont {T.}~\bibnamefont {Ozawa}}\ and\ \bibinfo {author} {\bibfnamefont {B.}~\bibnamefont {Mera}},\ }\bibfield  {title} {\bibinfo {title} {Relations between topology and the quantum metric for chern insulators},\ }\href {https://doi.org/10.1103/PhysRevB.104.045103} {\bibfield  {journal} {\bibinfo  {journal} {Phys. Rev. B}\ }\textbf {\bibinfo {volume} {104}},\ \bibinfo {pages} {045103} (\bibinfo {year} {2021})}\BibitemShut {NoStop}%
\bibitem [{\citenamefont {Mera}\ and\ \citenamefont {Ozawa}(2021{\natexlab{a}})}]{PhysRevB.104.045104}%
  \BibitemOpen
  \bibfield  {author} {\bibinfo {author} {\bibfnamefont {B.}~\bibnamefont {Mera}}\ and\ \bibinfo {author} {\bibfnamefont {T.}~\bibnamefont {Ozawa}},\ }\bibfield  {title} {\bibinfo {title} {K\"ahler geometry and chern insulators: Relations between topology and the quantum metric},\ }\href {https://doi.org/10.1103/PhysRevB.104.045104} {\bibfield  {journal} {\bibinfo  {journal} {Phys. Rev. B}\ }\textbf {\bibinfo {volume} {104}},\ \bibinfo {pages} {045104} (\bibinfo {year} {2021}{\natexlab{a}})}\BibitemShut {NoStop}%
\bibitem [{\citenamefont {Mera}\ and\ \citenamefont {Ozawa}(2021{\natexlab{b}})}]{PhysRevB.104.115160}%
  \BibitemOpen
  \bibfield  {author} {\bibinfo {author} {\bibfnamefont {B.}~\bibnamefont {Mera}}\ and\ \bibinfo {author} {\bibfnamefont {T.}~\bibnamefont {Ozawa}},\ }\bibfield  {title} {\bibinfo {title} {Engineering geometrically flat chern bands with fubini-study k\"ahler structure},\ }\href {https://doi.org/10.1103/PhysRevB.104.115160} {\bibfield  {journal} {\bibinfo  {journal} {Phys. Rev. B}\ }\textbf {\bibinfo {volume} {104}},\ \bibinfo {pages} {115160} (\bibinfo {year} {2021}{\natexlab{b}})}\BibitemShut {NoStop}%
\bibitem [{\citenamefont {Ledwith}\ \emph {et~al.}(2023)\citenamefont {Ledwith}, \citenamefont {Vishwanath},\ and\ \citenamefont {Parker}}]{PhysRevB.108.205144}%
  \BibitemOpen
  \bibfield  {author} {\bibinfo {author} {\bibfnamefont {P.~J.}\ \bibnamefont {Ledwith}}, \bibinfo {author} {\bibfnamefont {A.}~\bibnamefont {Vishwanath}},\ and\ \bibinfo {author} {\bibfnamefont {D.~E.}\ \bibnamefont {Parker}},\ }\bibfield  {title} {\bibinfo {title} {Vortexability: A unifying criterion for ideal fractional chern insulators},\ }\href {https://doi.org/10.1103/PhysRevB.108.205144} {\bibfield  {journal} {\bibinfo  {journal} {Phys. Rev. B}\ }\textbf {\bibinfo {volume} {108}},\ \bibinfo {pages} {205144} (\bibinfo {year} {2023})}\BibitemShut {NoStop}%
\bibitem [{\citenamefont {Fujimoto}\ \emph {et~al.}(2024)\citenamefont {Fujimoto}, \citenamefont {Parker}, \citenamefont {Dong}, \citenamefont {Khalaf}, \citenamefont {Vishwanath},\ and\ \citenamefont {Ledwith}}]{fujimoto2024higher}%
  \BibitemOpen
  \bibfield  {author} {\bibinfo {author} {\bibfnamefont {M.}~\bibnamefont {Fujimoto}}, \bibinfo {author} {\bibfnamefont {D.~E.}\ \bibnamefont {Parker}}, \bibinfo {author} {\bibfnamefont {J.}~\bibnamefont {Dong}}, \bibinfo {author} {\bibfnamefont {E.}~\bibnamefont {Khalaf}}, \bibinfo {author} {\bibfnamefont {A.}~\bibnamefont {Vishwanath}},\ and\ \bibinfo {author} {\bibfnamefont {P.}~\bibnamefont {Ledwith}},\ }\bibfield  {title} {\bibinfo {title} {Higher vortexability: zero field realization of higher landau levels},\ }\href {https://arxiv.org/abs/2403.00856} {\bibfield  {journal} {\bibinfo  {journal} {arXiv:2403.00856}\ } (\bibinfo {year} {2024})}\BibitemShut {NoStop}%
\bibitem [{\citenamefont {Wan}\ \emph {et~al.}(2023)\citenamefont {Wan}, \citenamefont {Sarkar}, \citenamefont {Lin},\ and\ \citenamefont {Sun}}]{PhysRevLett.130.216401}%
  \BibitemOpen
  \bibfield  {author} {\bibinfo {author} {\bibfnamefont {X.}~\bibnamefont {Wan}}, \bibinfo {author} {\bibfnamefont {S.}~\bibnamefont {Sarkar}}, \bibinfo {author} {\bibfnamefont {S.-Z.}\ \bibnamefont {Lin}},\ and\ \bibinfo {author} {\bibfnamefont {K.}~\bibnamefont {Sun}},\ }\bibfield  {title} {\bibinfo {title} {Topological exact flat bands in two-dimensional materials under periodic strain},\ }\href {https://doi.org/10.1103/PhysRevLett.130.216401} {\bibfield  {journal} {\bibinfo  {journal} {Phys. Rev. Lett.}\ }\textbf {\bibinfo {volume} {130}},\ \bibinfo {pages} {216401} (\bibinfo {year} {2023})}\BibitemShut {NoStop}%
\bibitem [{\citenamefont {Sarkar}\ \emph {et~al.}(2023)\citenamefont {Sarkar}, \citenamefont {Wan}, \citenamefont {Lin},\ and\ \citenamefont {Sun}}]{sarkar2023symmetry}%
  \BibitemOpen
  \bibfield  {author} {\bibinfo {author} {\bibfnamefont {S.}~\bibnamefont {Sarkar}}, \bibinfo {author} {\bibfnamefont {X.}~\bibnamefont {Wan}}, \bibinfo {author} {\bibfnamefont {S.-Z.}\ \bibnamefont {Lin}},\ and\ \bibinfo {author} {\bibfnamefont {K.}~\bibnamefont {Sun}},\ }\bibfield  {title} {\bibinfo {title} {Symmetry-based classification of exact flat bands in single and bilayer moir\'e systems},\ }\href {https://arxiv.org/abs/2310.02218} {\bibfield  {journal} {\bibinfo  {journal} {arXiv:2310.02218}\ } (\bibinfo {year} {2023})}\BibitemShut {NoStop}%
\bibitem [{\citenamefont {Lee}\ \emph {et~al.}(2019)\citenamefont {Lee}, \citenamefont {Khalaf}, \citenamefont {Liu}, \citenamefont {Liu}, \citenamefont {Hao}, \citenamefont {Kim},\ and\ \citenamefont {Vishwanath}}]{lee2019theory}%
  \BibitemOpen
  \bibfield  {author} {\bibinfo {author} {\bibfnamefont {J.~Y.}\ \bibnamefont {Lee}}, \bibinfo {author} {\bibfnamefont {E.}~\bibnamefont {Khalaf}}, \bibinfo {author} {\bibfnamefont {S.}~\bibnamefont {Liu}}, \bibinfo {author} {\bibfnamefont {X.}~\bibnamefont {Liu}}, \bibinfo {author} {\bibfnamefont {Z.}~\bibnamefont {Hao}}, \bibinfo {author} {\bibfnamefont {P.}~\bibnamefont {Kim}},\ and\ \bibinfo {author} {\bibfnamefont {A.}~\bibnamefont {Vishwanath}},\ }\bibfield  {title} {\bibinfo {title} {Theory of correlated insulating behaviour and spin-triplet superconductivity in twisted double bilayer graphene},\ }\href {http://dx.doi.org/10.1038/s41467-019-12981-1} {\bibfield  {journal} {\bibinfo  {journal} {Nat. Commun.}\ }\textbf {\bibinfo {volume} {10}},\ \bibinfo {pages} {5333} (\bibinfo {year} {2019})}\BibitemShut {NoStop}%
\bibitem [{\citenamefont {Su}\ and\ \citenamefont {Lin}(2020)}]{PhysRevLett.125.226401}%
  \BibitemOpen
  \bibfield  {author} {\bibinfo {author} {\bibfnamefont {Y.}~\bibnamefont {Su}}\ and\ \bibinfo {author} {\bibfnamefont {S.-Z.}\ \bibnamefont {Lin}},\ }\bibfield  {title} {\bibinfo {title} {Current-induced reversal of anomalous hall conductance in twisted bilayer graphene},\ }\href {https://doi.org/10.1103/PhysRevLett.125.226401} {\bibfield  {journal} {\bibinfo  {journal} {Phys. Rev. Lett.}\ }\textbf {\bibinfo {volume} {125}},\ \bibinfo {pages} {226401} (\bibinfo {year} {2020})}\BibitemShut {NoStop}%
\bibitem [{\citenamefont {Han}\ \emph {et~al.}(2023)\citenamefont {Han}, \citenamefont {Lu}, \citenamefont {Scuri}, \citenamefont {Sung}, \citenamefont {Wang}, \citenamefont {Han}, \citenamefont {Watanabe}, \citenamefont {Taniguchi}, \citenamefont {Fu}, \citenamefont {Park} \emph {et~al.}}]{han2023orbital}%
  \BibitemOpen
  \bibfield  {author} {\bibinfo {author} {\bibfnamefont {T.}~\bibnamefont {Han}}, \bibinfo {author} {\bibfnamefont {Z.}~\bibnamefont {Lu}}, \bibinfo {author} {\bibfnamefont {G.}~\bibnamefont {Scuri}}, \bibinfo {author} {\bibfnamefont {J.}~\bibnamefont {Sung}}, \bibinfo {author} {\bibfnamefont {J.}~\bibnamefont {Wang}}, \bibinfo {author} {\bibfnamefont {T.}~\bibnamefont {Han}}, \bibinfo {author} {\bibfnamefont {K.}~\bibnamefont {Watanabe}}, \bibinfo {author} {\bibfnamefont {T.}~\bibnamefont {Taniguchi}}, \bibinfo {author} {\bibfnamefont {L.}~\bibnamefont {Fu}}, \bibinfo {author} {\bibfnamefont {H.}~\bibnamefont {Park}}, \emph {et~al.},\ }\bibfield  {title} {\bibinfo {title} {Orbital multiferroicity in pentalayer rhombohedral graphene},\ }\href {https://www.nature.com/articles/s41586-023-06572-w} {\bibfield  {journal} {\bibinfo  {journal} {Nature}\ }\textbf {\bibinfo {volume} {623}},\ \bibinfo {pages} {41} (\bibinfo {year} {2023})}\BibitemShut {NoStop}%
\bibitem [{\citenamefont {Haldane}(1991)}]{PhysRevLett.67.937}%
  \BibitemOpen
  \bibfield  {author} {\bibinfo {author} {\bibfnamefont {F.~D.~M.}\ \bibnamefont {Haldane}},\ }\bibfield  {title} {\bibinfo {title} {``fractional statistics'' in arbitrary dimensions: A generalization of the pauli principle},\ }\href {https://doi.org/10.1103/PhysRevLett.67.937} {\bibfield  {journal} {\bibinfo  {journal} {Phys. Rev. Lett.}\ }\textbf {\bibinfo {volume} {67}},\ \bibinfo {pages} {937} (\bibinfo {year} {1991})}\BibitemShut {NoStop}%
\bibitem [{\citenamefont {Johnson}\ and\ \citenamefont {Canright}(1994)}]{PhysRevB.49.2947}%
  \BibitemOpen
  \bibfield  {author} {\bibinfo {author} {\bibfnamefont {M.~D.}\ \bibnamefont {Johnson}}\ and\ \bibinfo {author} {\bibfnamefont {G.~S.}\ \bibnamefont {Canright}},\ }\bibfield  {title} {\bibinfo {title} {Haldane fractional statistics in the fractional quantum hall effect},\ }\href {https://doi.org/10.1103/PhysRevB.49.2947} {\bibfield  {journal} {\bibinfo  {journal} {Phys. Rev. B}\ }\textbf {\bibinfo {volume} {49}},\ \bibinfo {pages} {2947} (\bibinfo {year} {1994})}\BibitemShut {NoStop}%
\bibitem [{\citenamefont {Li}\ \emph {et~al.}(2024)\citenamefont {Li}, \citenamefont {Su}, \citenamefont {Kim}, \citenamefont {Kee}, \citenamefont {Sun},\ and\ \citenamefont {Lin}}]{li2024contrasting}%
  \BibitemOpen
  \bibfield  {author} {\bibinfo {author} {\bibfnamefont {H.}~\bibnamefont {Li}}, \bibinfo {author} {\bibfnamefont {Y.}~\bibnamefont {Su}}, \bibinfo {author} {\bibfnamefont {Y.~B.}\ \bibnamefont {Kim}}, \bibinfo {author} {\bibfnamefont {H.-Y.}\ \bibnamefont {Kee}}, \bibinfo {author} {\bibfnamefont {K.}~\bibnamefont {Sun}},\ and\ \bibinfo {author} {\bibfnamefont {S.-Z.}\ \bibnamefont {Lin}},\ }\bibfield  {title} {\bibinfo {title} {Contrasting twisted bilayer graphene and transition metal dichalcogenides for fractional chern insulators: an emergent gauge picture},\ }\href {https://arxiv.org/abs/2402.02251} {\bibfield  {journal} {\bibinfo  {journal} {arXiv:2402.02251}\ } (\bibinfo {year} {2024})}\BibitemShut {NoStop}%
\bibitem [{\citenamefont {Abouelkomsan}\ \emph {et~al.}(2023)\citenamefont {Abouelkomsan}, \citenamefont {Yang},\ and\ \citenamefont {Bergholtz}}]{PhysRevResearch.5.L012015}%
  \BibitemOpen
  \bibfield  {author} {\bibinfo {author} {\bibfnamefont {A.}~\bibnamefont {Abouelkomsan}}, \bibinfo {author} {\bibfnamefont {K.}~\bibnamefont {Yang}},\ and\ \bibinfo {author} {\bibfnamefont {E.~J.}\ \bibnamefont {Bergholtz}},\ }\bibfield  {title} {\bibinfo {title} {Quantum metric induced phases in moir\'e materials},\ }\href {https://doi.org/10.1103/PhysRevResearch.5.L012015} {\bibfield  {journal} {\bibinfo  {journal} {Phys. Rev. Res.}\ }\textbf {\bibinfo {volume} {5}},\ \bibinfo {pages} {L012015} (\bibinfo {year} {2023})}\BibitemShut {NoStop}%
\end{thebibliography}%


\begin{thebibliography}{5}%
\makeatletter
\providecommand \@ifxundefined [1]{%
 \@ifx{#1\undefined}
}%
\providecommand \@ifnum [1]{%
 \ifnum #1\expandafter \@firstoftwo
 \else \expandafter \@secondoftwo
 \fi
}%
\providecommand \@ifx [1]{%
 \ifx #1\expandafter \@firstoftwo
 \else \expandafter \@secondoftwo
 \fi
}%
\providecommand \natexlab [1]{#1}%
\providecommand \enquote  [1]{``#1''}%
\providecommand \bibnamefont  [1]{#1}%
\providecommand \bibfnamefont [1]{#1}%
\providecommand \citenamefont [1]{#1}%
\providecommand \href@noop [0]{\@secondoftwo}%
\providecommand \href [0]{\begingroup \@sanitize@url \@href}%
\providecommand \@href[1]{\@@startlink{#1}\@@href}%
\providecommand \@@href[1]{\endgroup#1\@@endlink}%
\providecommand \@sanitize@url [0]{\catcode `\\12\catcode `\$12\catcode `\&12\catcode `\#12\catcode `\^12\catcode `\_12\catcode `\%12\relax}%
\providecommand \@@startlink[1]{}%
\providecommand \@@endlink[0]{}%
\providecommand \url  [0]{\begingroup\@sanitize@url \@url }%
\providecommand \@url [1]{\endgroup\@href {#1}{\urlprefix }}%
\providecommand \urlprefix  [0]{URL }%
\providecommand \Eprint [0]{\href }%
\providecommand \doibase [0]{https://doi.org/}%
\providecommand \selectlanguage [0]{\@gobble}%
\providecommand \bibinfo  [0]{\@secondoftwo}%
\providecommand \bibfield  [0]{\@secondoftwo}%
\providecommand \translation [1]{[#1]}%
\providecommand \BibitemOpen [0]{}%
\providecommand \bibitemStop [0]{}%
\providecommand \bibitemNoStop [0]{.\EOS\space}%
\providecommand \EOS [0]{\spacefactor3000\relax}%
\providecommand \BibitemShut  [1]{\csname bibitem#1\endcsname}%
\let\auto@bib@innerbib\@empty
\bibitem [{\citenamefont {Wan}\ \emph {et~al.}(2023)\citenamefont {Wan}, \citenamefont {Sarkar}, \citenamefont {Lin},\ and\ \citenamefont {Sun}}]{PhysRevLett.130.216401}%
  \BibitemOpen
  \bibfield  {author} {\bibinfo {author} {\bibfnamefont {X.}~\bibnamefont {Wan}}, \bibinfo {author} {\bibfnamefont {S.}~\bibnamefont {Sarkar}}, \bibinfo {author} {\bibfnamefont {S.-Z.}\ \bibnamefont {Lin}},\ and\ \bibinfo {author} {\bibfnamefont {K.}~\bibnamefont {Sun}},\ }\href {https://doi.org/10.1103/PhysRevLett.130.216401} {\bibfield  {journal} {\bibinfo  {journal} {Phys. Rev. Lett.}\ }\textbf {\bibinfo {volume} {130}},\ \bibinfo {pages} {216401} (\bibinfo {year} {2023})}\BibitemShut {NoStop}%
\bibitem [{\citenamefont {Li}\ \emph {et~al.}(2024)\citenamefont {Li}, \citenamefont {Su}, \citenamefont {Kim}, \citenamefont {Kee}, \citenamefont {Sun},\ and\ \citenamefont {Lin}}]{li2024contrasting}%
  \BibitemOpen
  \bibfield  {author} {\bibinfo {author} {\bibfnamefont {H.}~\bibnamefont {Li}}, \bibinfo {author} {\bibfnamefont {Y.}~\bibnamefont {Su}}, \bibinfo {author} {\bibfnamefont {Y.~B.}\ \bibnamefont {Kim}}, \bibinfo {author} {\bibfnamefont {H.-Y.}\ \bibnamefont {Kee}}, \bibinfo {author} {\bibfnamefont {K.}~\bibnamefont {Sun}},\ and\ \bibinfo {author} {\bibfnamefont {S.-Z.}\ \bibnamefont {Lin}},\ }\href {https://arxiv.org/abs/2402.02251} {\bibfield  {journal} {\bibinfo  {journal} {arXiv:2402.02251}\ } (\bibinfo {year} {2024})}\BibitemShut {NoStop}%
\bibitem [{\citenamefont {Bergholtz}\ and\ \citenamefont {Karlhede}(2008)}]{PhysRevB.77.155308}%
  \BibitemOpen
  \bibfield  {author} {\bibinfo {author} {\bibfnamefont {E.~J.}\ \bibnamefont {Bergholtz}}\ and\ \bibinfo {author} {\bibfnamefont {A.}~\bibnamefont {Karlhede}},\ }\href {https://doi.org/10.1103/PhysRevB.77.155308} {\bibfield  {journal} {\bibinfo  {journal} {Phys. Rev. B}\ }\textbf {\bibinfo {volume} {77}},\ \bibinfo {pages} {155308} (\bibinfo {year} {2008})}\BibitemShut {NoStop}%
\bibitem [{\citenamefont {Regnault}\ and\ \citenamefont {Bernevig}(2011)}]{PhysRevX.1.021014}%
  \BibitemOpen
  \bibfield  {author} {\bibinfo {author} {\bibfnamefont {N.}~\bibnamefont {Regnault}}\ and\ \bibinfo {author} {\bibfnamefont {B.~A.}\ \bibnamefont {Bernevig}},\ }\href {https://doi.org/10.1103/PhysRevX.1.021014} {\bibfield  {journal} {\bibinfo  {journal} {Phys. Rev. X}\ }\textbf {\bibinfo {volume} {1}},\ \bibinfo {pages} {021014} (\bibinfo {year} {2011})}\BibitemShut {NoStop}%
\bibitem [{\citenamefont {Bernevig}\ and\ \citenamefont {Haldane}(2008)}]{PhysRevLett.100.246802}%
  \BibitemOpen
  \bibfield  {author} {\bibinfo {author} {\bibfnamefont {B.~A.}\ \bibnamefont {Bernevig}}\ and\ \bibinfo {author} {\bibfnamefont {F.~D.~M.}\ \bibnamefont {Haldane}},\ }\href {https://doi.org/10.1103/PhysRevLett.100.246802} {\bibfield  {journal} {\bibinfo  {journal} {Phys. Rev. Lett.}\ }\textbf {\bibinfo {volume} {100}},\ \bibinfo {pages} {246802} (\bibinfo {year} {2008})}\BibitemShut {NoStop}%
\end{thebibliography}%
 
\end{document}


\title{Supplementary information for  ``Quantum metric induced quantum Hall conductance inversion and reentrant transition in fractional Chern insulators''}
\author{Ang-Kun Wu}
\affiliation{Theoretical Division, T-4, Los Alamos National Laboratory, Los Alamos, New Mexico 87545, USA}
\author{Siddhartha Sarkar}
\affiliation{Department of Physics, University of Michigan, Ann Arbor, Michigan, 48109, USA}
\author{Xiaohan Wan}
\affiliation{Department of Physics, University of Michigan, Ann Arbor, Michigan, 48109, USA}
\author{Kai Sun}
\email{sunkai@umich.edu}
\affiliation{Department of Physics, University of Michigan, Ann Arbor, Michigan, 48109, USA}
\author{Shi-Zeng Lin}
\email{szl@lanl.gov}
\affiliation{Theoretical Division, T-4 and CNLS, Los Alamos National Laboratory,
Los Alamos, New Mexico 87545, USA}
\affiliation{Center for Integrated Nanotechnology, Los Alamos National Laboratory,
Los Alamos, New Mexico 87545, USA}

\maketitle

\tableofcontents

\section{Quantum geometry of single-particle flat bands}

Based on Ref. \cite{PhysRevLett.130.216401}, we know that there are magic values for the strain strength characterized by a dimensionless parameter $\Tilde{\alpha}$, where exact topological flat bands emerge in the chiral limit $c_0=0$. These topological flat bands are sublattice polarized and their wave functions are holomorphic. In this work, we focus on the first $2$ magic parameters $\Tilde{\alpha}_1=0.78943,\ \Tilde{\alpha}_2=2.1325$, where the standard deviation of the Berry curvature in the momentum space for $\Tilde{\alpha}_1$ is $\mathrm{std}(F_{xy})=0.281$, which is one order of magnitude larger than that for the second $\Tilde{\alpha}_2$, $\mathrm{std}(F_{xy})=0.027$. 

\begin{figure}[t!]
\begin{center}
\includegraphics[width = 0.7\textwidth]{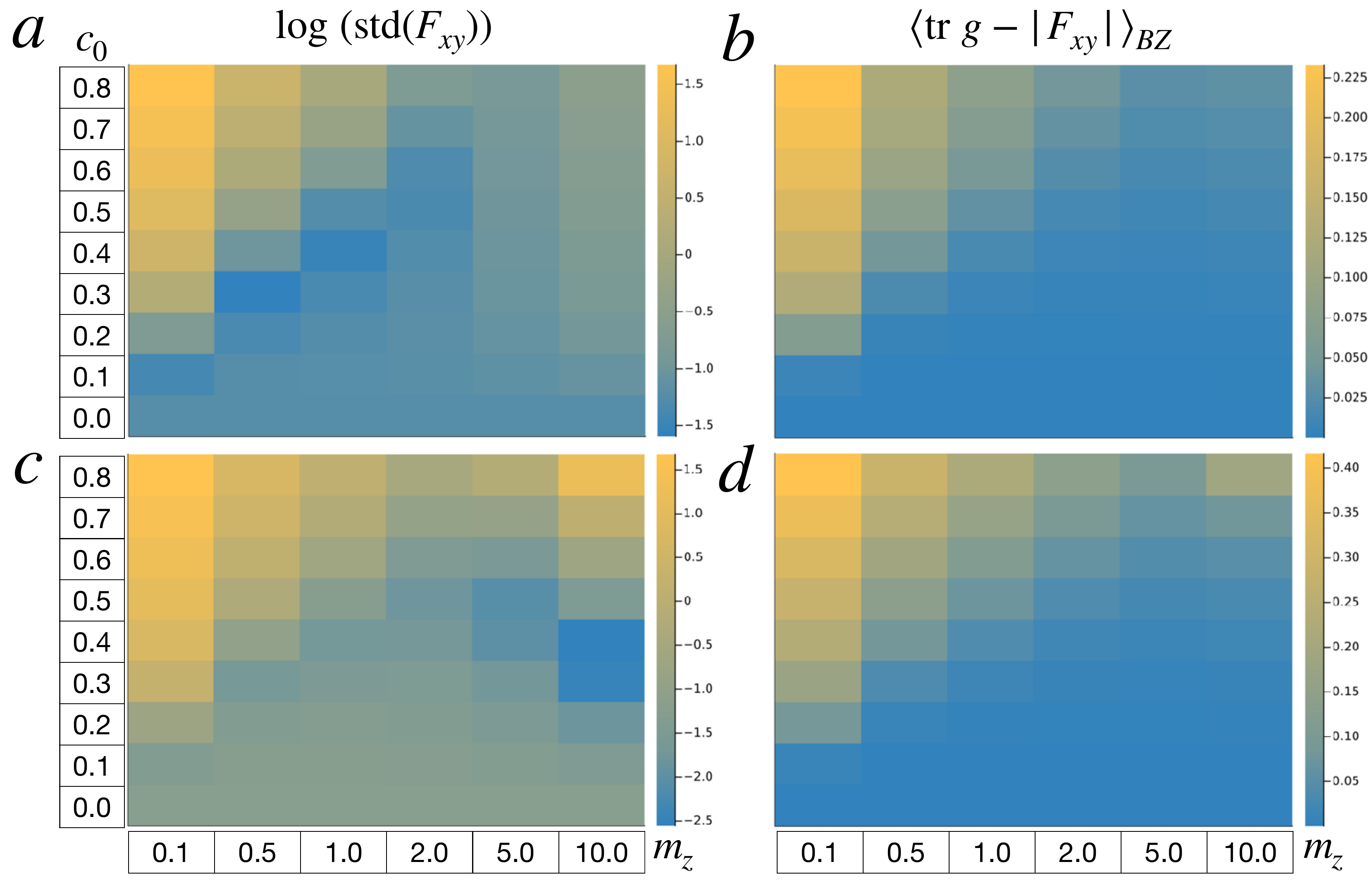}
\caption{Standard deviation of Berry curvature in the momentum space $\mathrm{std}(F_{xy})$ (a,c) and averaged deviation of the trace condition $\mathrm{tr} g-|F_{xy}|$ (b, d) for the first magic parameter $\Tilde{\alpha}_1=0.78943$. (a, b) are plots for the lower flat band with Chern number $C=-1$ and (c, d) are plots for the higher Chern band $C=1$. }
\label{fig:a1geometry}
\end{center}
\end{figure}

\begin{figure}[h!]
\begin{center}
\includegraphics[width = 0.7\textwidth]{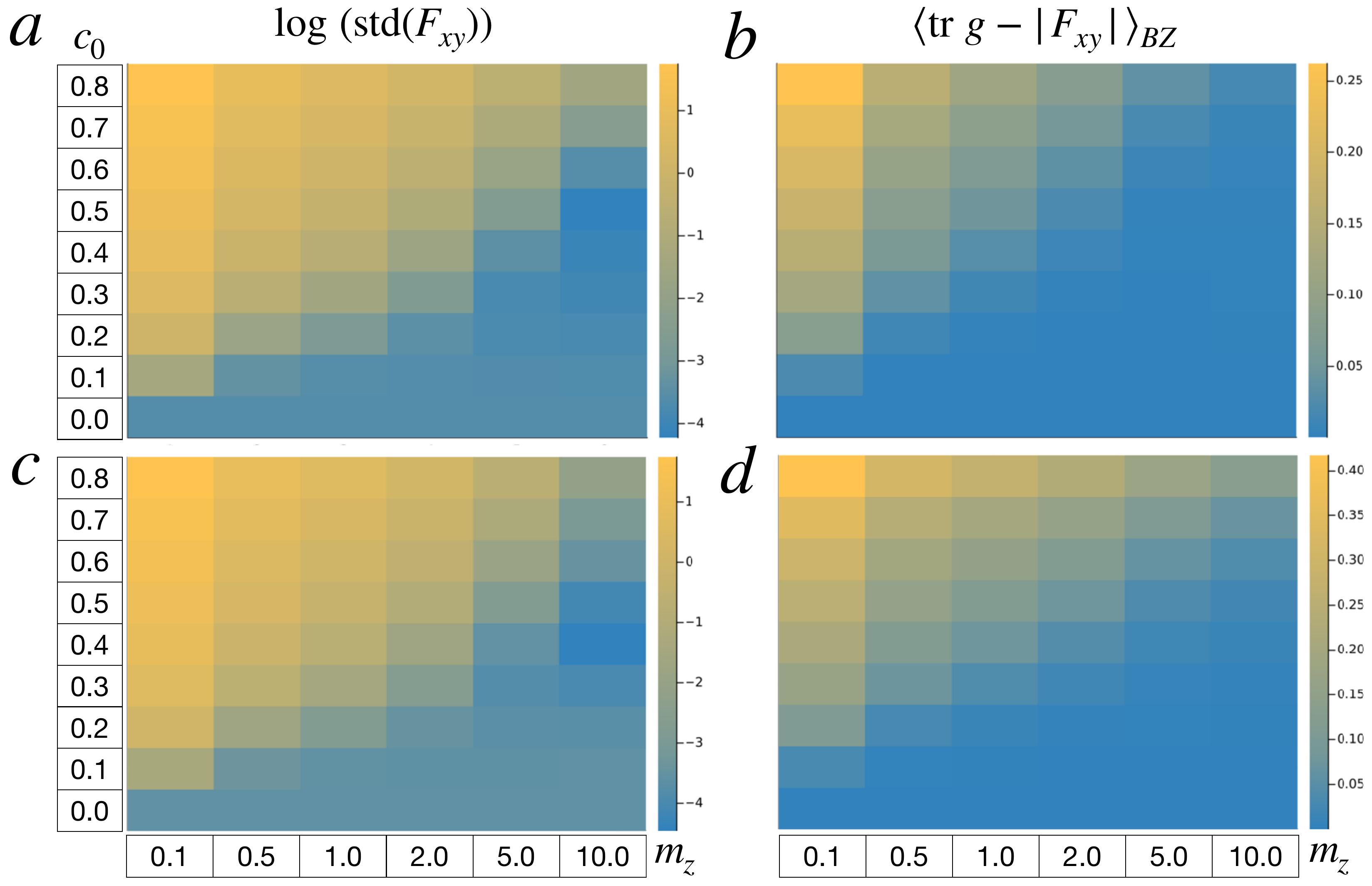}
\caption{Standard deviation of Berry curvature in the momentum space $\mathrm{std}(F_{xy})$ (a,c) and averaged deviation of the trace condition $\mathrm{tr} g-|F_{xy}|$ (b, d) for the second magic parameter $\Tilde{\alpha}_2=2.1325$. (a, b) are plots for the lower flat band with Chern number $C=-1$ while (c, d) are plots for the higher Chern band $C=1$. }
\label{fig:a2geometry}
\end{center}
\end{figure}

In Figs. \ref{fig:a1geometry} and \ref{fig:a2geometry}, we show the Berry curvature and quantum geometry of the two topological flat bands for various $c_0,\ m_z$ away from the chiral limit. In all parameter ranges shown in Figs. \ref{fig:a1geometry} and \ref{fig:a2geometry}, the two flat bands near charge neutrality are topological with $C=\pm 1$ (larger $c_0$ will make the two bands hybridize with other remote bands). In the chiral limit $c_0=0$, the trace condition $\mathrm{tr} g-|F_{xy}|=0$ is satisfied in the whole Brillouin zone (BZ). Generally, moving away from the chiral limit (increasing $c_0$) violates the trace condition more strongly. However, the Berry curvature distribution is different for the two magic parameters. For the second magic parameter $\Tilde{\alpha}_2$ in Fig. \ref{fig:a2geometry} (\textbf{a,c}), since the Berry curvature is almost uniform in the chiral limit, increasing $c_0$ increases $\mathrm{std}(F_{xy})$. In contrast, for $\Tilde{\alpha}_1=0.78943$, $\mathrm{std}(F_{xy})$ first decreases with $c_0$ and then increases, especially for a small $m_z$, as shown in Fig. \ref{fig:a1geometry} (\textbf{a,c}).

As $c_0$ breaks the sublattice symmetry for the $C=\pm 1$ flat bands, the top $C=1$ flat band has a larger quantum metric $\int \mathrm{d}\mathbf{k} \; \mathrm{tr}[g(\mathbf{k})]$ than the lower flat band $C=-1$ at $c_0>0$. In Figs. \ref{fig:a1geometry} and \ref{fig:a2geometry}, the $C=1$ band has a larger $\langle \mathrm{tr} g-|F_{xy}|\rangle_{BZ}$ than the $C=-1$ band because $\langle |F_{xy}|\rangle_{BZ}$ is a constant, equal to $2\pi/N$ with $N$ the number of momentum points. The situation reverses for $c_0<0$, where the higher $C=1$ band has a smaller quantum metric $\int \mathrm{d}\mathbf{k} \; \mathrm{tr}[g(\mathbf{k})]$ than that of the lower $C=-1$ band. As explained in the main text, this gives rise to inverted FCI states as a result of the shift of the band dispersion caused by the Fock contribution.

\newpage
\section{Momentum grid for exact diagonalization}
\label{SISec:grid}

\begin{figure}[h!]
\begin{center}
\includegraphics[width = 0.8 \textwidth]{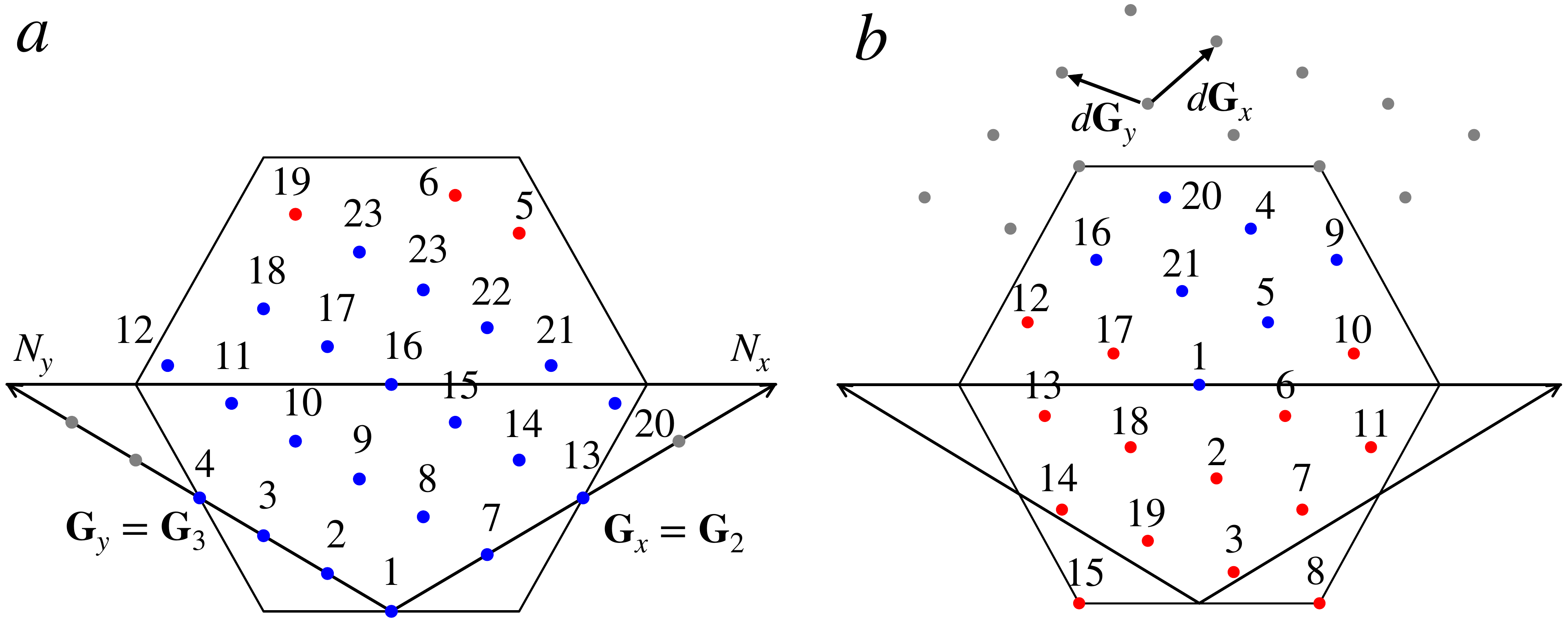}
\caption{Schematic view of momentum grid used in the ED calculations. (a) The regular momentum grid choice by dividing the reciprocal lattice vectors $\mathbf{G}_2,\ \mathbf{G}_3$ into $N_x=4,\ N_y=6$ equal divisions. (b) A more uniform momentum grid choice with $N=21$ with the $K,\ K'$ and $\Gamma$ points included. Here the unit vectors of the grid are $d\mathbf{G}_x=(5G_2,G_3)/21,d\mathbf{G}_y=(G_2,-4G_3)/21$. The blue and red circles are $\mathbf{k}$ points constrained to the first Brillouin zone (BZ). The red circles are obtained by folding the gray points back to first BZ.}
\label{fig:kgrid}
\end{center}
\end{figure}

The choices of momentum grids can help reduce the finite-size effect for the exact diagonalization. In principle, the more uniformly we cover over the first Moir'e Brillouin zone (BZ), the better numerical results we should have for finite-size systems. When it comes to FCI states, the choices of the momentum grid might change the total momentum sectors where the degenerate ground states appear. In addition, we need to select high-symmetry points to treat the competing CDW states on an equal footing, i.e., the $\mathbf{K}$, $\mathbf{K'}$, and $\Gamma$ points for the hexagonal lattice at $v=1/3$ filling. In the main text, we used two sets of momentum grids as shown in Fig. \ref{fig:kgrid}. The first is a uniform division of reciprocal vector $\mathbf{G}_x=\mathbf{G}_2,\mathbf{G}_y=\mathbf{G}_3$ by integers $N_x,\ N_y$ with a total of $\mathbf{k}$ points $N=N_xN_y$. The advantage of this choice is that we can conveniently add twist phases along $\mathbf{G}_x,\mathbf{G}_y$ directions and compute the many-body Chern number. The disadvantage is that for some system sizes, i.e., $N=21=3\times 7,N=27=3\times 9$, we have much denser $\mathbf{k}$ points in one direction than in the other, which amplifies the finite-size effect. For ED calculations of systems with two bands, $N=21$ is the largest system size we can reach. We use the second momentum grid shown in Fig. \ref{fig:kgrid}(b), where we first fix the $\mathbf{K},\ \mathbf{K'}$, and $\Gamma$ points, then optimize the uniformity for $N=21$ and obtain $d\mathbf{G}_x=(5\mathbf{G}_2,\mathbf{G}_3)/21,d\mathbf{G}_y=(\mathbf{G}_2,-4\mathbf{G}_3)/21$ (the momentum index choice is the same as in \cite{li2024contrasting}).

\newpage
\section{More characterizations of the inverted FCI states}

\begin{figure}[h!]
\begin{center}
\includegraphics[width = 1.0 \textwidth]{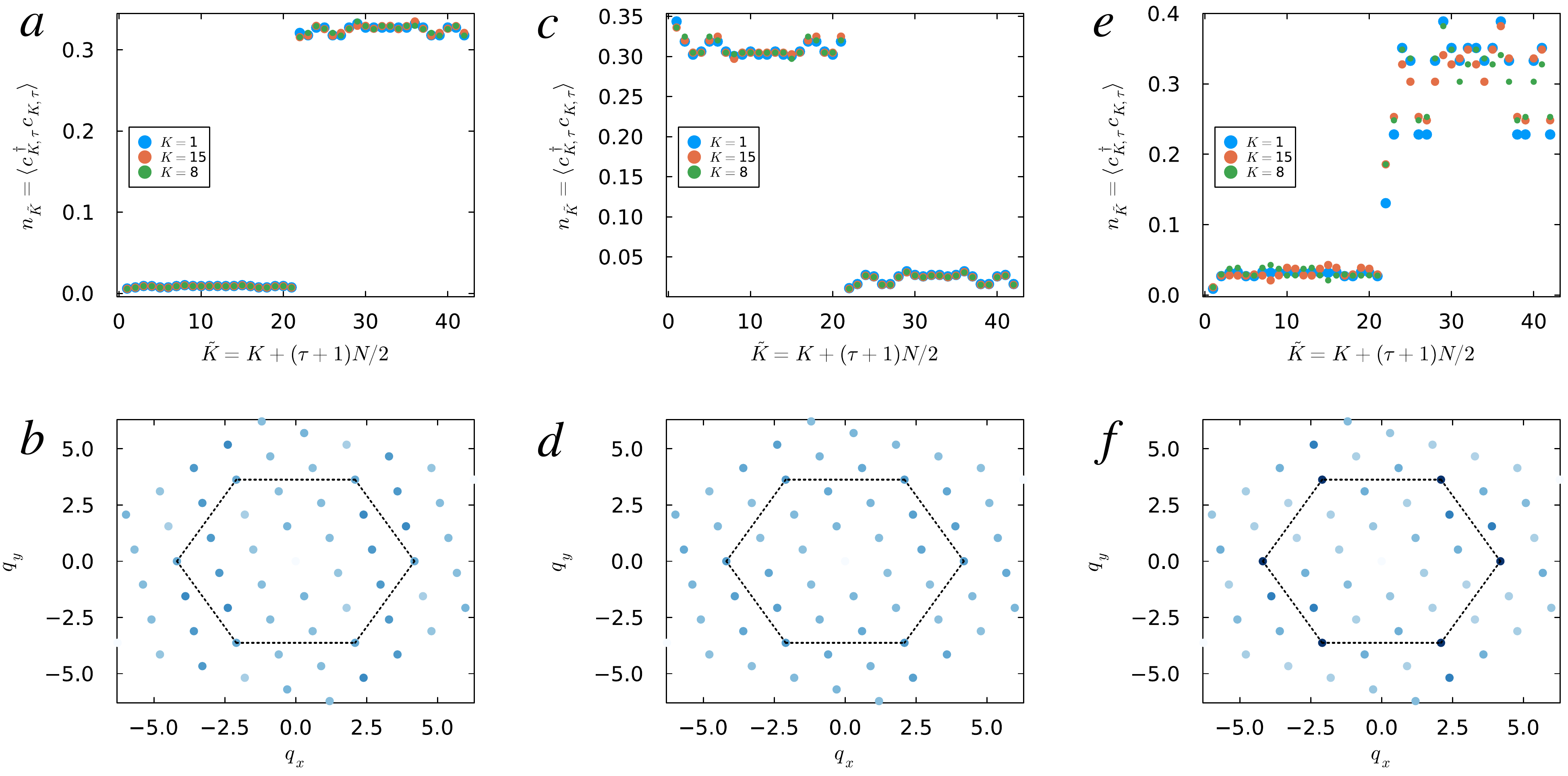}
\caption{Density distribution in the momentum space $n_{\tilde{K}}$ (first row) and density-density correlation $|\tilde{C}(\mathbf{q})|$ (second row) of the lowest $9$ states in Fig. 1 in the main text. The parameters used in the ED calculation are $c_0=-0.25,\ m_z=1, N=21$ with the $\mathbf{k}$ grid defined in Fig. \ref{fig:kgrid}b in SI Sec. \ref{SISec:grid}. (a, b) $n_{\tilde{K}}$ and $|\tilde{C}(\mathbf{q})|$ for the lowest $3$ states. (c, d) The second lowest $3$ states. (e, f) The third lowest $3$ states.  The density-density correlation is calculated for the lowest energy state of each set of $3$ states. Darker (lighter) color indicates higher (lower) intensity.
The dotted hexagon marks the first BZ.}
\label{fig:FCImore}
\end{center}
\end{figure}

To further show that the lowest $3$ FCI states are predominantly composed of the single particle states of the higher flat topological band, the most straightforward way is to look at the density occupancy $n_{\tilde{K}}=\langle c^\dagger_{K,\tau} c_{K,\tau} \rangle$, where $K$ denotes the momentum grid index, $\tau=\pm 1$ is the band index, and $\tilde{K}=K+(\tau+1)N/2$ is the extended index to include both bands. From the first row of Fig. \ref{fig:FCImore}, it is clear that the lowest $3$ states have uniform occupancy at the higher band, while the second lowest $3$ states uniformly occupy the lower band. Interestingly, the third lowest $3$ states, which give a CDW-like behavior, mostly occupy the higher band.

Another quantity to check is the density-density correlation function
\begin{equation}
    \begin{split}
        C(\mathbf{q})=\langle \rho(\mathbf{q})\rho(-\mathbf{q})\rangle-\langle \rho(\mathbf{q})\rangle \langle \rho(-\mathbf{q})\rangle,
    \end{split}
\end{equation}
where $\rho(\mathbf{q})=\sum_{\mathbf{k},\tau,\tau'}\lambda_{\tau,\tau',\mathbf{q}}(\mathbf{k})c^\dagger_{\mathbf{k},\tau} c_{\mathbf{k+q},\tau'}$ and $\lambda_{\tau,\tau',\mathbf{q}}=\langle u_{\tau,\mathbf{k}}|u_{\tau',\mathbf{k}}\rangle$. 
For convenience, we also examine the density-density correlation for bare density $n_\mathbf{q}$, which describes the correlation on a computational basis.
\begin{equation}
    \tilde{C}(\mathbf{q})=\langle n_\mathbf{q} n_{\mathbf{-q}}\rangle -\langle n_\mathbf{q}\rangle \langle n_{\mathbf{-q}}\rangle,\quad n_\mathbf{q}\equiv \sum_\mathbf{k} c^{\dagger}_{\mathbf{k+q}}c_\mathbf{k}.
\end{equation}
In the FCI phase, this correlation should be uniform as a function of the momentum $\mathbf{q}$, while in the CDW phase \cite{PhysRevB.77.155308}, this correlation is peaked at some special momenta $\mathbf{q}$. In the second row of Fig. \ref{fig:FCImore}, the low energy states show very uniform correlation (subplot \textbf{b, d}, indicating they are FCI states), while the high energy state ($7$th state) shows peaks in the density-density correlation. Combined with the particle entanglement spectrum counting in the main text, this state is identified as a CDW state.

\textit{Details for the calculation of $\tilde{C}(\mathbf{q})$:} Since the total momentum $K$ is conserved for a given eigenstate $|\Psi\rangle$, $\langle\Psi|c^\dagger_{\mathbf{k+q}}c_\mathbf{k}|\Psi\rangle=\delta_{0\mathbf{q}}$, $\langle n_\mathbf{q}\rangle=\langle \Psi|n_\mathbf{q}|\Psi\rangle=\delta_{0\mathbf{q}}\langle \sum_\mathbf{k} c^\dagger_\mathbf{k}c_\mathbf{k}\rangle=\delta_{0\mathbf{q}}Q$. For the first term,
\begin{equation}
    \begin{split}
        \langle n_\mathbf{q} n_{\mathbf{-q}}\rangle &=\langle \sum_\mathbf{k_1,k_2} c^\dagger_\mathbf{k_1+q}c_\mathbf{k_1}c^\dagger_\mathbf{k_2-q}c_\mathbf{k_2}\rangle.
    \end{split}
\end{equation}
Due to the antisymmetry of the fermionic wavefunction, we also need to be careful with signs, i.e., $c^\dagger_2 c^\dagger_1|\Omega\rangle =|0011\rangle =-c^\dagger_1 c^\dagger_2|\Omega\rangle$, where the momentum vector is ordered into 1D indices. 

(1) When $\mathbf{q}=0$
\begin{equation}
    \begin{split}
        \langle n_\mathbf{q=0} n_{\mathbf{-q=0}}\rangle &=\langle 2\sum_{k_1<k_2} c^\dagger_{k_1}c_{k_1}c^\dagger_{k_2}c_{k_2}+\sum_{k}c^\dagger_k c_k\rangle=\langle 2\sum_{k_1<k_2} c^\dagger_{k_2}c^\dagger_{k_1}c_{k_1}c_{k_2}\rangle +Q.
    \end{split}
\end{equation}

(2) When $\mathbf{q}\ne 0, \mathbf{k_1+q}\ne k_1, \mathbf{k_2-q}\ne k_2$, in this case, $k_1\ne k_2$,  
\begin{equation}
    \begin{split}
        \langle n_\mathbf{q} n_{\mathbf{-q}}\rangle &=\langle \sum_{k_1 \ne k_2,\mathbf{k_2-q\ne k_1}} c^\dagger_{k_1+q}c_{k_1}c^\dagger_{k_2-q}c_{k_2}+\sum_{\mathbf{k_2-q= k_1}} c^\dagger_{k_2}c_{k_1}c^\dagger_{k_1}c_{k_2}\rangle\\
        &=\langle \sum_{k_1 \ne k_2,\mathbf{k_2-q\ne k_1}} c^\dagger_{k_1+q}c_{k_1}c^\dagger_{k_2-q}c_{k_2}+\sum_{k_2} c^\dagger_{k_2} c_{k_2}-\sum_{k_2,\mathbf{k_2-q= k_1}} c^\dagger_{k_2}c^\dagger_{k_1}c_{k_1}c_{k_2}\rangle\\
        &=\langle \sum_{k_1 \ne k_2,\mathbf{k_2-q\ne k_1}} c^\dagger_{k_1+q}c_{k_1}c^\dagger_{k_2-q}c_{k_2}-\sum_{k_2,\mathbf{k_2-q= k_1}} c^\dagger_{k_2}c^\dagger_{k_1}c_{k_1}c_{k_2}\rangle+Q.
    \end{split}
\end{equation}
Note that the sign of the operator $c^\dagger_{k_1+q}c_{k_1}c^\dagger_{k_2-q}c_{k_2}$ acting on a basis depends on the order of $k_1+q,k_1,k_2-q,k_2$. Just like in the generation of the many-body Hamiltonian, there is a sign $(-1)^{\mu(k_1,k_2)}$ where $\mu(k_1,k_2)$ counts the number of occupancy in the current basis.

\newpage
\section{Inverted FCI at other fillings}

\begin{figure}[h!]
\begin{center}
\includegraphics[width = 1.0 \textwidth]{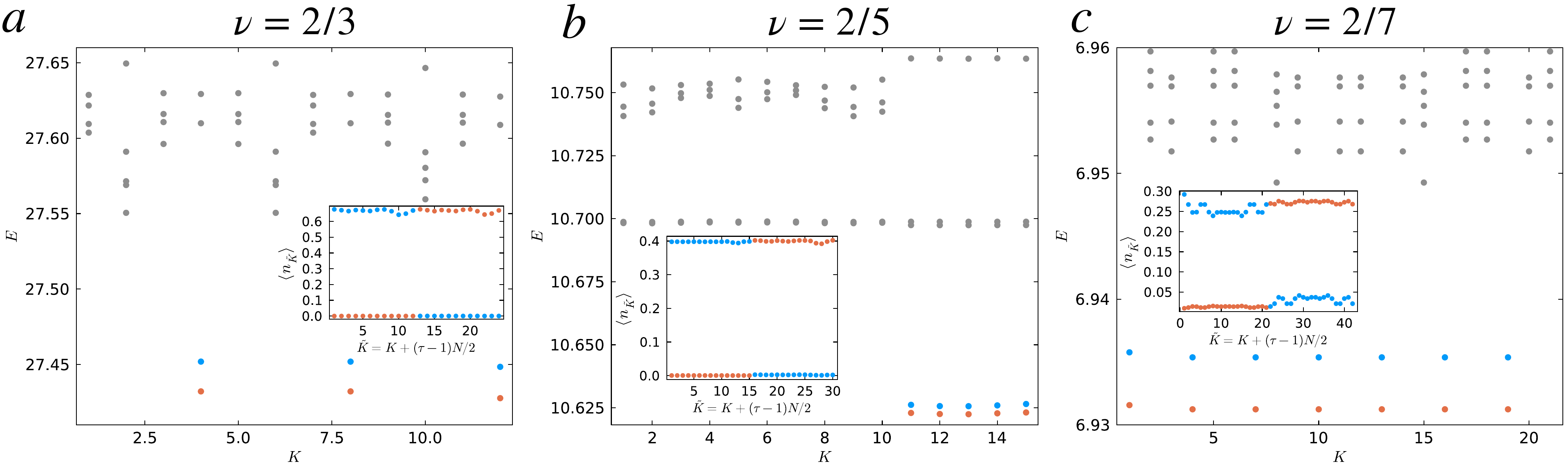}
\caption{Many-body energy spectrum and density distributions (insets). The parameters are $c_0=-0.1,\ m_z=1,\tilde{\alpha}=2.1325$. (a) $\nu=2/3$ with system size $N_x=3,N_y=4$. (b) $\nu=2/5$ with system size $N_x=3,N_y=5$. (c) $\nu=2/7$ with system size $N=21$ and grid scheme defined in Fig. \ref{fig:kgrid}b in SI Sec. \ref{SISec:grid}. Note that $c_0=-0.25$ since the effect of interband hybridization is somehow stronger at this filling, see Sec. \ref{sec:interband}.}
\label{fig:fillings}
\end{center}
\end{figure}

The inverted FCI states also appear at other fillings, e.g., $\nu=2/3,\ 2/5,\ 2/7$, corresponding to many-body Chern number $C_\mathrm{MB}=\nu$. (Note that the lower single-particle band has Chern number -1). In Fig. \ref{fig:fillings}, we show the many-body energy spectrum. At filling $\nu=2/q$ with an odd $q$ (it is easier to stabilize the FCI states with the numerator $2$), there are two sets of FCI states with $q$-fold degeneracy. When we consider the $q$ degenerate states and look at the average density, for all fillings, the lowest $q$ states have a uniform density $n_{\tilde{K}}\sim \nu$ in the higher single-particle band, while the second lowest $q$ states uniformly occupy the lower single-particle band. The facts that (1) the second lowest $q$ states  mostly occupy lower band and (2) their density is uniform indicate that these excited states are also FCI states but from the lower band.
Therefore, the many-body Chern number of the ground FCI state is inverted in comparison to the single-particle Chern number of the lower band.

\newpage
\section{Calculation of the many-body Chern number}

We use twisted boundary condition (TBC) to scan all the momentum points in the Brillouin zone. A $2\pi$ twist phase along one direction means move the $\mathbf{k}$ grid by one unit $1/N_x,1/N_y$ of the reciprocal lattice $\mathbf{G}_2,\mathbf{G}_3$. When inserting flux by TBC to calculate the many-body Chern number, the single-particle basis is constantly changed, as is the many-body basis. For instance, given the twist phases $\phi_x,\ \phi_y$,
\begin{equation}
\begin{split}
    |\Psi(\phi_x,\phi_y)\rangle &= a|1010\cdots\rangle+b|0101\cdots\rangle +\cdots,\\
    |\Psi(\phi_x+d\phi,\phi_y)\rangle &= a'|1'0'1'0'\cdots\rangle+b'|0'1'0'1'\cdots\rangle +\cdots,
\end{split}
\end{equation}
where the occupancy basis are different, given by the single-particle orbitals (wavefunctions). Thus, the proper calculation of the wavefunction overlap should be
\begin{equation}
\begin{split}
    \langle \Psi(\phi_x,\phi_y)|\Psi(\phi_x+d\phi,\phi_y)\rangle&= a^*a' \langle 1010\cdots |1'0'1'0'\cdots\rangle+b^*b'\langle 0101\cdots|0'1'0'1'\cdots\rangle +\cdots.
\end{split}
\end{equation}
Here, the inner product of the basis $\langle 1010\cdots |1'0'1'0'\cdots\rangle \ne 1$, given by
\begin{equation}
\begin{split}
    \langle 1010\cdots |1'0'1'0'\cdots\rangle = \langle u_1(\phi_x,\phi_y)|u_1'(\phi_x+d\phi,\phi_y)\rangle \langle u_3(\phi_x,\phi_y)|u_3'(\phi_x+d\phi,\phi_y)\rangle \cdots.
\end{split}
\end{equation}
For the generic twist phase $\Delta\phi$, the quantum geometry is involved in both 
\begin{align}
    \langle 1010\cdots |1'0'1'0'\cdots\rangle\ne 0,
\end{align}
\begin{align}
    \langle 1010\cdots |0'1'1'0'\cdots\rangle\ne 0.
\end{align}
However, with a relatively large number of electrons, $Q$, we can approximate orthogonal bases, i.e.,
\begin{align}
    \langle 1'0'1'0'\cdots |0110\cdots\rangle =0.
\end{align}
This introduces the single-particle geometry to the inner product of many-body states for the calculation of the many-body Chern number.
Using this proper approach for overlapping different ground states, instead of calculating the Berry curvature, we can divide the twist phases into plaquettes and use loop overlaps (inner products) of the ground state at plaquette corners to obtain the Berry phase for each plaquette. Summing over all plaquettes gives the total Berry phase.

\newpage
\section{Particle Entanglement Spectrum (PES) and excitation counting}

Any quantum many-body states can be Schmidt decomposed into two parts $A,B$ (not necessarily spatial orbitals)
\begin{equation}
    |\Psi\rangle = \sum_i e^{-\xi_i/2} |\Psi_i^A\rangle \otimes |\Psi_i^B\rangle,
\end{equation}
where $\langle \Psi_i^\alpha|\Psi_j^\beta\rangle = \delta_{i,j}\delta_{\alpha\beta}$ with $\alpha,\beta\in [A,B].$ Here $e^{-\xi_i}$ and $|\Psi_i^\alpha\rangle$ are the eigenvalues and eigenstates of the reduced density matrix 
\begin{equation}
    \rho_A = \mathrm{Tr}_B \rho, \quad \rho = |\Psi\rangle \langle \Psi|.
\end{equation}
Normally, the regions $A, B$ are defined as spatially separated orbitals. For instance, if there are 2 particles in 4 orbitals, the basis consists of 6 states
\begin{equation}
    |1,1,0,0\rangle, |1,0,1,0\rangle,|1,0,0,1\rangle,|0,1,1,0\rangle,|0,1,0,1\rangle,|0,0,1,1\rangle,
\end{equation}
where $1,0$ denote occupied and empty orbitals and the four digits represent $4$ orbitals. In the spatial partition of orbitals, $A$ could be the first $2$ orbitals while $B$ will be the last $2$
\begin{equation}
\begin{split}
    &|1,1\rangle_A\otimes |0,0\rangle_B,|1,0\rangle_A\otimes |1,0\rangle_B,|1,0\rangle_A\otimes |0,1\rangle_B\\
    &|0,1\rangle_A\otimes |1,0\rangle_B,|0,1\rangle_A\otimes |0,1\rangle_B,|0,0\rangle_A\otimes |1,1\rangle_B.
\end{split}
\end{equation}
If we represent a state by a matrix $|\Psi\rangle = \sum_{ij} M_{ij} |\Psi_i\rangle^A \otimes |\Psi_j\rangle^B$ (just a reshape of the state coefficients), then
\begin{equation}
    \begin{split}
        \rho&=|\Psi\rangle\langle \Psi |=\sum_{ijkl} M_{ij} M^*_{kl} |\Psi_i\rangle_A \langle \Psi_k|_A \otimes |\Psi_j\rangle_B \langle \Psi_l|_B\\
        \rho_A &= \mathrm{Tr}_B \rho=\sum_{ik} (\sum_j M_{ij} M^*_{kj})|\Psi_i\rangle_A \langle \Psi_k|_A = \sum_{ik} \{\rho_A\}_{ik} |\Psi_i\rangle_A \langle \Psi_k|_A,
    \end{split}
\end{equation}
where $M_{kj}^*=\{M^\dagger\}_{jk}, \rho_A = M M^\dagger$. Thus, instead of constructing a huge matrix $\rho$ and then do the trace, we can first reshape the state coefficients into a matrix $M$ and then use $\rho_A=MM^\dagger$.

For the case where we separate $A,B$ by particle indices instead of orbital indices, we can rewrite the previous basis 
$|1,2\rangle,|1,3\rangle,|1,4\rangle$, $|2,3\rangle,|2,4\rangle,|3,4\rangle$, where the $2$ digits represent the particle locations and $1,2,3,4$ represent $4$ particle indices. When $A$ represents the first particle (in the order) and $B$ is the second, we have the decomposition.
\begin{equation}
\begin{split}
    &|1\rangle_A\otimes |2\rangle_B,|1\rangle_A\otimes |3\rangle_B,|1\rangle_A\otimes |4\rangle_B,|2\rangle_A\otimes |3\rangle_B,|2\rangle_A\otimes |4\rangle_B,|3\rangle_A\otimes |4\rangle_B.
\end{split}
\end{equation}
This partition guarantees that there is always $1$ particle in each region. Compared to the previous partition, the dimensionality of regions $A,B$ becomes $3$ (compared to $4$ previously). Scaling with systems sizes and particle numbers, however, shows that the dimensionality is larger for the particle partition, with the Hilbert space dimension $D_H=C^L_{Q/2}$ with $C^n_k$ being the combination formula, as compared to the orbital partition $2^{L/2}$ because $L \ge Q$ for fermions. Note that for the orbital partition, each orbital is either occupied or unoccupied, which is independent. However, for the particle partition, particle $1$ can only occupy orbital $1,\cdots, L-Q+1$, particle $2$ can only occupy $2,\cdots, L-Q+2$, ......, and particle $Q$ can only occupy $Q,\cdots, L$. 
Thus, the dimensionality of the particle partition configurations is larger than the orbital partition. For a given state $|1,2\rangle$ it can be separated into two partitions 
$$|1\rangle_A\otimes |2\rangle_B, \quad |2\rangle_A\otimes |1\rangle_B,$$
where the dimensionality for each partition is $D_H=C^L_{Q_{A|B}}$. Besides, particles are identical and we should consider particles without ordering as particles $1,2$. 

To properly separate electrons based on particle partition, it is better to expand in real space coordinates via Slater determinant
\begin{equation}
\begin{split}
    |abcd\rangle &= c^\dagger_d c^\dagger_c c^\dagger_b c^\dagger_a |0\rangle \\
    \to \langle \mathbf{x}|abcd\rangle &= \phi_a(x_1)\phi_b(x_2)\otimes\phi_c(x_3) \phi_d(x_4)-\phi_b(x_1)\phi_a(x_2)\otimes\phi_c(x_3) \phi_d(x_4)\\
    &-\phi_a(x_1)\phi_c(x_2)\otimes\phi_b(x_3) \phi_d(x_4)+\phi_c(x_1)\phi_a(x_2)\otimes\phi_b(x_3) \phi_d(x_4)+\cdots
\end{split}
\end{equation}
where there are $Q!$ terms in the Slater determinant. On the other hand, it can be reorganized with decomposition based on particle indices,
\begin{equation}
\begin{split}
    \langle \mathbf{x}|abcd\rangle &= [\phi_a(x_1)\phi_b(x_2)-\phi_b(x_1)\phi_a(x_2)]\otimes[\phi_c(x_3) \phi_d(x_4)-\phi_d(x_3) \phi_c(x_4)]\\
    &-[\phi_a(x_1)\phi_c(x_2)-\phi_c(x_1)\phi_a(x_2)]\otimes[\phi_b(x_3) \phi_d(x_4)-\phi_b(x_3) \phi_b(x_4)]+\cdots
\end{split}
\end{equation}
Here, there are also $C^Q_{Q_A} (Q_A!)(Q_B!)$ terms, which are equal to $Q!$. The decomposition helps us go back to the Fock space within particle partition
\begin{equation}
\begin{split}
    |abcd\rangle &= |ab\rangle \otimes|cd\rangle-|ac\rangle\otimes|bd\rangle+\cdots \\
    &= \sum_{\sigma} \mathrm{sgn}(\sigma) |\sigma_A\rangle \otimes |\sigma_B\rangle = \sum_{ij} M_{ij} \psi^A_i \otimes \psi^B_j,
\end{split}
\end{equation}
where $\mathrm{sgn}(\sigma)$ is the signature of permutation, i.e., $\sigma=(acbd)$, and $\sigma_A$ is first $Q_A$ part of $\sigma$ while $\sigma_B$ constitutes the remaining part.

To accurately obtain the counting in the entanglement spectrum of the $\nu=1/3$ fractional Chern insulating states, the density matrix should be an average of the $3$ degenerate ground states
\begin{equation}
    \rho = \frac{1}{3}\sum_{i=1}^3 |\Psi_i\rangle \langle \Psi_i|.
\end{equation}
Note that construction of the density matrix using a single ground state or an average of 3 ground states do not work. The count of $\nu=1/3$ is given by Sec. VII in \cite{PhysRevX.1.021014} 
\begin{equation}
    \frac{(N-2Q_A-1)!}{(Q_A!)(N-3Q_A!)}N,
\end{equation}
where $N=N_xN_y$ is the total number of momentum points. This combinatorial counting is obtained by counting the $(1,3)-$ admissible states (separating balls with blocks) with the periodic boundary condition \cite{PhysRevLett.100.246802,PhysRevX.1.021014}.

\newpage

\section{Self-consistent Hartree-Fock approximation}\label{SIsec:HF}

The total Hamiltonian with single-particle dispersion and interaction is
\begin{equation}
    \begin{split}
        H = \sum_{\mathbf{k},\tau} (\epsilon_{\mathbf{k},\tau}-\mu) c^\dagger_{\mathbf{k},\tau} c_{\mathbf{k},\tau}+\frac{1}{2A} \sum_{\mathbf{q}} \rho(\mathbf{q})V(\mathbf{q})\rho(-\mathbf{q}),
    \end{split}
\end{equation}
where $A$ is the volume (area) of the BZ, and $\tau$ is the band index. To begin with, let us consider the chiral limit, where we do not have interband hybridization. The density operator is
\begin{equation}
    \rho(\mathbf{q})=\sum_{\mathbf{k,k'},\tau} \langle \psi_{\mathbf{k},\tau} |e^{i\mathbf{q\cdot r}} |\psi_{\mathbf{k'},\tau} \rangle c^\dagger_{\mathbf{k},\tau} c_{\mathbf{k'},\tau}=\sum_{\mathbf{k},\tau}\lambda_{\tau,\mathbf{q}}(\mathbf{k}) c^\dagger_{\mathbf{k},\tau} c_{\mathbf{k+q},\tau},
\end{equation}
where $\psi_{\mathbf{k},\tau}(\mathbf{r})=e^{i\mathbf{k\cdot r}}u_{\tau,\mathbf{k}}(\mathbf{r})=\frac{1}{\sqrt{\Omega}} \sum_\mathbf{G} e^{i\mathbf{(G+k)\cdot r}}u_{\tau,\mathbf{k}}(\mathbf{G})$ is the Bloch state and the form factor $\lambda_{\tau,\mathbf{q}}(\mathbf{k})=\langle u_{\tau,\mathbf{k}}|u_{\tau,\mathbf{k+q}}\rangle$. The total Hamiltonian in the chiral limit becomes
\begin{equation}
    H = \sum_{\mathbf{k},\tau} (\epsilon_{\mathbf{k},\tau}-\mu) c^\dagger_{\mathbf{k},\tau} c_{\mathbf{k},\tau}+\frac{1}{2A} \sum_{\mathbf{k,k',q},\tau,\tau'} \lambda_{\tau,\mathbf{q}}(\mathbf{k}) c^\dagger_{\mathbf{k},\tau} c_{\mathbf{k+q},\tau}V(\mathbf{q})\lambda_{\tau',-\mathbf{q}}(\mathbf{k'}) c^\dagger_{\mathbf{k'},\tau'} c_{\mathbf{k-q},\tau'}.
\end{equation}
We use the Hartree-Fock mean field approach to calculate the ground state energy. The Hartree contribution to the total energy is
\begin{equation}
    \begin{split}
        &\frac{1}{2A} \sum_{\mathbf{k,k',q},\tau,\tau'} \lambda_{\tau,\mathbf{q}}(\mathbf{k}) \overbracket{c^\dagger_{\mathbf{k},\tau} c_{\mathbf{k+q},\tau}}V(\mathbf{q})\lambda_{\tau',-\mathbf{q}}(\mathbf{k'}) \overbracket{c^\dagger_{\mathbf{k'},\tau'} c_{\mathbf{k'-q},\tau'}}\\
        &=\frac{1}{2A} \sum_{\mathbf{k,k',q},\tau,\tau'} \lambda_{\tau,\mathbf{q}}(\mathbf{k}) \langle c^\dagger_{\mathbf{k},\tau} c_{\mathbf{k+q},\tau} \rangle V(\mathbf{q})\lambda_{\tau',-\mathbf{q}}(\mathbf{k'}) c^\dagger_{\mathbf{k'},\tau'} c_{\mathbf{k'-q},\tau'}\\
        &+\frac{1}{2A} \sum_{\mathbf{k,k',q},\tau,\tau'} \lambda_{\tau,\mathbf{q}}(\mathbf{k}) c^\dagger_{\mathbf{k},\tau} c_{\mathbf{k+q},\tau} V(\mathbf{q})\lambda_{\tau',-\mathbf{q}}(\mathbf{k'}) \langle c^\dagger_{\mathbf{k'},\tau'} c_{\mathbf{k'-q},\tau'}\rangle \\
        &=\frac{1}{A}\sum_{\mathbf{q}\in \mathbf{G}} \sum_{\mathbf{k},\tau} \lambda_{\tau,\mathbf{q}}(\mathbf{k}) c^\dagger_{\mathbf{k},\tau} c_{\mathbf{k},\tau} V(\mathbf{q}) \sum_{\mathbf{k'},\tau'} \lambda_{\tau',-\mathbf{q}} (\mathbf{k'})f(E_{\mathbf{k'},\tau'}),
    \end{split}
\end{equation}
where $f(E)$ is the Fermi distribution, and here we restrict to zero temperature. The Fock contribution is
\begin{equation}
    \begin{split}
        &\frac{1}{2A} \sum_{\mathbf{k,k',q},\tau,\tau'} \lambda_{\tau,\mathbf{q}}(\mathbf{k}) \overbracket{c^\dagger_{\mathbf{k},\tau} \underbracket{c_{\mathbf{k+q},\tau} V(\mathbf{q})\lambda_{\tau',-\mathbf{q}}(\mathbf{k'}) c^\dagger_{\mathbf{k'},\tau'}} c_{\mathbf{k-q},\tau'}}\\
        &=-\frac{1}{2A} \sum_{\mathbf{k,k',q},\tau,\tau'} \lambda_{\tau,\mathbf{q}}(\mathbf{k}) \langle c^\dagger_{\mathbf{k},\tau} c_{\mathbf{k'-q},\tau'} \rangle V(\mathbf{q})\lambda_{\tau',-\mathbf{q}}(\mathbf{k'}) c^\dagger_{\mathbf{k'},\tau'} c_{\mathbf{k+q},\tau}\\
        &-\frac{1}{2A} \sum_{\mathbf{k,k',q},\tau,\tau'} \lambda_{\tau,\mathbf{q}}(\mathbf{k}) c^\dagger_{\mathbf{k},\tau} c_{\mathbf{k'-q},\tau'} V(\mathbf{q})\lambda_{\tau',-\mathbf{q}}(\mathbf{k'}) \langle c^\dagger_{\mathbf{k'},\tau'} c_{\mathbf{k+q},\tau} \rangle\\
        &= -\frac{1}{A} \sum_\mathbf{q}\sum_{\mathbf{k,k'=k+q}} \lambda_{\tau,\mathbf{q}}(\mathbf{k})c^\dagger_{\mathbf{k},\tau} c_{\mathbf{k'-q},\tau} V(\mathbf{q})\lambda_{\tau,-\mathbf{q}}(\mathbf{k'})f(E_{\mathbf{k'},\tau'}).
    \end{split}
\end{equation}
Combining the two terms, we obtain the total energy
\begin{equation}
    \begin{split}
        E_{\mathbf{k},\tau} &= \epsilon_{\mathbf{k},\tau}-\mu+\frac{1}{A}\sum_{\mathbf{q}=n\mathbf{G}} \lambda_{\tau,\mathbf{q}}(\mathbf{k})V(\mathbf{q})\sum_{\mathbf{k'},\tau'} \lambda_{\tau',-\mathbf{q}}(\mathbf{k'})f(E_{\mathbf{k'},\tau'})\\
        &-\frac{1}{A}\sum_{\mathbf{q,k'}} \lambda_{\tau,\mathbf{q}}(\mathbf{k})V(\mathbf{q})\lambda_{\tau,-\mathbf{q}}(\mathbf{k'})f(E_{\mathbf{k'},\tau})
    \end{split}
\end{equation}

In a similar fashion, we can introduce interband hybridization in the density operator
\begin{equation}
    \rho(\mathbf{q})=\sum_{\mathbf{k,k'},\tau,\tau'} \langle \psi_{\mathbf{k},\tau}|e^{i\mathbf{q\cdot r}}|\psi_{\mathbf{k'},\tau'}\rangle c^\dagger_{\mathbf{k},\tau} c_{\mathbf{k+q},\tau'}=\sum_{\mathbf{k},\tau,\tau'} \lambda_{\tau,\tau',\mathbf{q}}(\mathbf{k}) c^\dagger_{\mathbf{k},\tau} c_{\mathbf{k+q},\tau'}.
\end{equation}

The self-consistent Hartree-Fock approximation is
\begin{equation}
    \begin{split}
        E_{\mathbf{k},\tau} &= \epsilon_{\mathbf{k},\tau}-\mu+\frac{1}{A}\sum_{\mathbf{q}=n\mathbf{G}}  \lambda_{\tau,\tau,\mathbf{q}}(\mathbf{k})V(\mathbf{q})\sum_{\mathbf{k'},\tau'} \lambda_{\tau',\tau',-\mathbf{q}}(\mathbf{k'})f(E_{\mathbf{k'},\tau'})\\
        &-\frac{1}{A}\sum_{\mathbf{q,k'=k+q}} \sum_{\tau'}\lambda_{\tau',\tau,\mathbf{q}}(\mathbf{k})V(\mathbf{q})\lambda_{\tau,\tau',-\mathbf{q}}(\mathbf{k'})f(E_{\mathbf{k'},\tau'}).
    \end{split}
\end{equation}
To have left-hand side $E_{\mathbf{k},\tau}$ for $c^\dagger_{\mathbf{k},\tau}c_{\mathbf{k},\tau}$, we need to constrain $\tau=\tau_2$ while for $\langle c^\dagger_{\mathbf{k},\tau_3} c_{\mathbf{k+q},\tau_4} \rangle=\delta_{\mathbf{q=0}}\delta_{\tau_3,\tau_4} f(E_{\mathbf{k},\tau_3})$ for the Hartree term. For the Fock term, they are $\mathbf{k'=k+q},\tau_2=\tau_3$ and $\langle c^\dagger_{\mathbf{k},\tau_1} c_{\mathbf{k'-q},\tau_4} \rangle=\delta_{\mathbf{k'=k+q}}\delta_{\tau_1,\tau_4} f(E_{\mathbf{k},\tau_1})$.


\textbf{Connect the Fock term to the quantum metric:} the Fock term contribution to a single band can be written as
\begin{equation}
\begin{split}
    \Delta_F(\mathbf{k},\tau)&\equiv -\frac{1}{A}\sum_{\mathbf{q,k'=k+q}} \lambda_{\tau,\mathbf{q}}(\mathbf{k})V(\mathbf{q})\lambda_{\tau,-\mathbf{q}}(\mathbf{k'})f(E_{\mathbf{k'},\tau})\\
    &\approx -\frac{1}{A}\sum_{\mathbf{q,k'=k+q}} V(q) (1-s^2_\tau(\mathbf{k,k+q}) f(E_{\mathbf{k},\tau}),
\end{split}
\end{equation}
where $\lambda_{\tau,-\mathbf{q}}(\mathbf{k+q})=\lambda_{\tau,\mathbf{q}}^*(\mathbf{k}).$ $s^2_\tau(\mathbf{k,k+q})$ is the Hilbert-Schmidt quantum distance of one band
\begin{equation}
    s^2_\tau(\mathbf{k,k+q})\equiv 1-|\langle u_{\tau,\mathbf{k}}|u_{\tau,\mathbf{k+q}}\rangle |^2=1-|\lambda_{\tau,\mathbf{q}}(\mathbf{k})|^2
\end{equation}
where the band index $\tau$ is suppressed. When $q$ is small,
\begin{equation}
    \langle u_{\mathbf{k}}|u_{\mathbf{k+q}}\rangle \approx 1+\langle u_{\mathbf{k}}|\partial_\mu u_{\mathbf{k}}\rangle q_\mu+\frac{1}{2!}\langle u_{\mathbf{k}}|\partial_{\mu,\nu} u_{\mathbf{k}}\rangle q_\mu q_\nu,
\end{equation}
where $\mu,\nu=x,y$ denote the component of $\mathbf{q}$ in 2D. With $\langle u_\mathbf{k}|\partial_\mu u_\mathbf{k}\rangle+ \langle \partial_\mu u_\mathbf{k}| u_\mathbf{k}\rangle=0$, it is straightforward to show that
\begin{equation}
\begin{split}
    s^2(\mathbf{k,k+q})&\approx \mathrm{Re}[\langle \partial_\mu u_\mathbf{k}|(1-|u_\mathbf{k}\rangle\langle u_\mathbf{k}|)|\partial_\nu u_\mathbf{k}\rangle]q_\mu q_\nu\\
    &=g_{\mu\nu} q_\mu q_\nu,
\end{split}
\end{equation}
where $\eta_{\mu\nu}(\mathbf{k})\equiv \langle \partial_\mu u_\mathbf{k}|(1-|u_\mathbf{k}\rangle\langle u_\mathbf{k}|)|\partial_\nu u_\mathbf{k}\rangle$ is the quantum geometric tensor and $g_{\mu\nu}(\mathbf{k})\equiv \mathrm{Re} [\eta_{\mu\nu}(\mathbf{k})]$ is the Fubini-Study metric.

For fast decay $V(q)$, the Fock term can be approximated as
\begin{equation}
    \begin{split}
        \Delta_F(\mathbf{k},\tau) &\approx -\frac{1}{A} \sum_{\mathbf{q},q<q_c} (1-g_{\mu\nu}(\mathbf{k},\tau)q_\mu q_\nu)V(q)f(E_{\mathbf{k},\tau})\\
        &=-\frac{1}{A} \int \mathrm{d}\mathbf{q} (1-g_{\mu\nu}(\mathbf{k},\tau)q_\mu q_\nu)V(q)f(E_{\mathbf{k},\tau})\\
        &=O(\mathbf{k},\tau)+\frac{\pi}{A} \mathrm{tr}[g_{\mu\nu}(\mathbf{k},\tau)]f(E_{\mathbf{k},\tau})\int \mathrm{d}q q^3 V(q)\\
        &= O(\mathbf{k},\tau)+\frac{\pi V_\mathrm{int}}{A} \mathrm{tr}[g_{\mu\nu}(\mathbf{k},\tau)]f(E_{\mathbf{k},\tau})
    \end{split}
\end{equation}
where $O(\mathbf{k},\tau) \equiv -\frac{1}{A} \int \mathrm{d}\mathbf{q} V(q)f(E_{\mathbf{k+q},\tau})\approx-\frac{2\pi}{A} \int  \mathrm{d}q V(q)f(E_{\mathbf{k},\tau})=-\frac{2\pi}{A} \int \mathrm{d}q V(q)=\mathcal{O}(1)$ and $V_\mathrm{int}\equiv\int \mathrm{d}q q^3 V(q)$. 
When $\mu\ne \nu$, the integral $\int \mathrm{d}\mathbf{q} g_{\mu\nu}(\mathbf{k},\tau)q_\mu q_\nu =\int \mathrm{d}q q g_{\mu\nu}(\mathbf{k},\tau)\int_0^{2\pi} \mathrm{d}\theta q\cos\theta q\sin\theta=0.$ When $\mu=\nu$,
\begin{equation}
    \int \mathrm{d}\mathbf{q} g_{\mu\mu}(\mathbf{k},\tau)q_\mu q_\mu=\int \mathrm{d}q q g_{\mu\mu}(\mathbf{k},\tau)\int_0^{2\pi} \mathrm{d}\theta q^2\cos^2\theta=\pi \int \mathrm{d}q q^3 g_{\mu\mu}(\mathbf{k},\tau).
\end{equation}

\newpage
\section{The role of interband hybridization}\label{sec:interband}

\begin{figure}[h!]
\begin{center}
\includegraphics[width = 0.8 \textwidth]{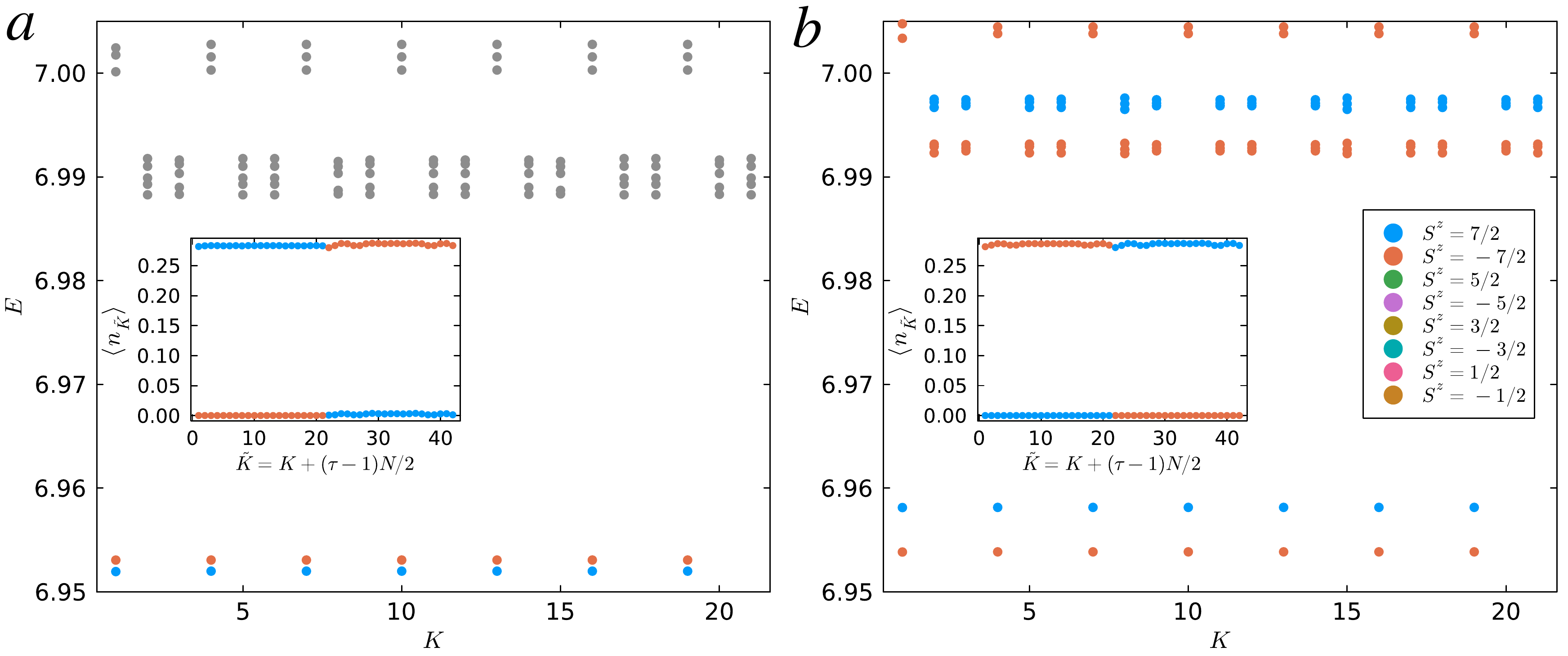}
\caption{Many-body energy spectrum and density distributions (insets) at $\nu=2/7$. The parameters are $c_0=-0.1,\ m_z=1,\ \tilde{\alpha}=2.1325$ with a system size $N=21$ and the grid scheme defined in Fig. \ref{fig:kgrid}b in SI Sec. \ref{SISec:grid}. (a) The spectrum with interband hybridization. (b) The spectrum without interband hybridization, where the electron polarization of topological bands is a good quantum number, $S^z\equiv \langle \sum_{\mathbf{k},\tau} \tau(\mathbf{k},\tau)\rangle/2$. The color scheme is the same as in Fig. 2 in the main text.}
\label{fig:interband}
\end{center}
\end{figure}

At the integer filling, $\nu=1$, electrons fully occupy the band with a smaller total trace of the quantum metric for a wide range of $c_0$ regardless of the interband hybridization according to Hartree-Fock calculations. However, when it comes to fractional fillings, the interband hybridization plays a more subtle role. For instance, at a generic fractional filling such as $\nu=1/3,\ 2/3,\ 2/5$, the inverted FCI states are observed at $c_0<0$ due to the quantum metric. This is not the case for $\nu=2/7$. The FCI states are not inverted when $c_0=-0.1$ [see Fig. \ref{fig:interband} (a)], but inverted for $c_0=0.1$ (not shown) states. To understand this exotic behavior, we turn off the interband hybridization. As shown in Fig. \ref{fig:interband} (b), the inverted FCI states now appear at $c_0=-0.1$, similar to $\nu=1/3$. The results indicate that interband hybridization plays an opposite role compared to the quantum metric in achieving the inverted FCI. When the difference in the quantum metric between the two bands becomes large for a strong deviation from the chiral limit, i.e. $c_0=-0.25$, the quantum metric dominates over the interband hybridization, and as a consequence, the inverted FCI states appear as shown in Fig. \ref{fig:fillings} (c). 

In addition to inverting the many-body Chern number of the FCI states compared to the expectation from partial filling of the single-particle band, the interband hybridization also affects the stability of the FCI states with respect to other competing states. The interband hybridization tends to mix two topological bands, and neutralize the band topology in the case with $C=\pm 1$ bands, which in turn destabilize the FCI states. For example, in the infinite $U$ limit, turning off the interband hybridization at $\nu=1/3$ causes the (inverted) FCI states to persist for much larger values of $|c_0|$ with a stronger deviation from the chiral limit.

\bibliography{SIrefs}